\documentclass[12pt,english,sort]{article}
\usepackage{mathptmx}

\usepackage[T1]{fontenc}
\usepackage[latin9]{inputenc}
\usepackage[a4paper]{geometry}
\usepackage{fancyhdr}
\usepackage{babel}
\usepackage{amsbsy}
\usepackage{amssymb}
\usepackage{graphicx}
\usepackage{setspace}
\usepackage{esint}
\usepackage[numbers]{natbib}
\usepackage{breakurl}

\makeatletter
\usepackage{amstext,amsbsy,amsopn,amsthm,amscd,amsfonts,amssymb}
\renewcommand\boldsymbol[1]{\pmb{#1}} 

\date{}

\renewcommand{\@openbib@code}{\setlength{\itemsep}{-1pt}}

\renewcommand{\subsectionmark}[1]{}

\newcommand{\begineq}[1]{\begin{equation}\label{#1}}
\newcommand{\fineq}{\end{equation}}
\newcommand{\Eq}[1]{Eq.\ (\ref{#1})}

\makeatother

\begin{document}
\begin{onehalfspace}
\global\long\def\refposT{\mathbf{X}}
\global\long\def\curposT{\mathbf{x}}
\global\long\def\mapping{\mathbf{\boldsymbol{\varphi}}}
\global\long\def\d{\mathrm{d}}
\global\long\def\indexunit{\hat{\mathbf{r}}_{0}^{\left(i\right)}}

\global\long\def\defgT{\mathbf{F}}
 \global\long\def\CGstrainT{\mathbf{C}}
 \global\long\def\CGleftstrainT{\mathbf{b}}
 \global\long\def\stretch{\lambda}

\global\long\def\EfieldT{\mathbf{E}}
\global\long\def\electricpotential{\phi}
\global\long\def\DfieldT{\mathbf{D}}
\global\long\def\polarizationT{\mathbf{P}}
\global\long\def\dipoleT{\mathbf{m}}

\global\long\def\Efield{E}
\global\long\def\Dfield{D}

\global\long\def\permittivity{\varepsilon}
\global\long\def\vacpermittivity{\varepsilon_{0}}
\global\long\def\susceptibility{\chi}

\global\long\def\EfieldTref{\mathbf{E}^{\left(0\right)}}
\global\long\def\DfieldTref{\mathbf{D}^{\left(0\right)}}
\global\long\def\polarizationTref{\mathbf{P}^{\left(0\right)}}

\global\long\def\EfieldRef{E^{\left(0\right)}}
\global\long\def\DfieldRef{D^{\left(0\right)}}

\global\long\def\energy{W}
\global\long\def\stressT{\boldsymbol{\sigma}}
\global\long\def\piolaT{\boldsymbol{\sigma}^{\left(0\right)}}
\global\long\def\mechanicalstressT{\boldsymbol{\sigma}_{m}}
\global\long\def\electricstressT{\boldsymbol{\pi}}
\global\long\def\maxwellstressT{\boldsymbol{\mathbf{\sigma}}_{v}}

\global\long\def\tractionT{\mathbf{t}}
\global\long\def\shear{\mu}
\global\long\def\chainspervolume{N}
\global\long\def\numberoflinks{n_{l}}
\global\long\def\numberofdipoles{n_{d}}

\global\long\def\linklen{l}
\global\long\def\langevin{\mathcal{L}}
\global\long\def\invlangevin{\mathcal{L}^{-1}}

\global\long\def\chainlencur{r}
\global\long\def\chainlenref{r_{0}}
\global\long\def\chainveccur{\mathbf{r}}
\global\long\def\chainvecref{\mathbf{r}_{0}}
\global\long\def\constant{\mathfrak{K}}

\global\long\def\rotation{\mathbf{R}}
\global\long\def\dipoledir{\hat{\boldsymbol{\xi}}}
\global\long\def\dipoledirref{\hat{\boldsymbol{\xi}}_{0}}

\end{onehalfspace}

\begin{onehalfspace}

\title{\vspace{-20 mm}Towards a physics-based modelling of the electro-mechanical
coupling in EAPs}
\end{onehalfspace}

\begin{onehalfspace}

\author{Noy Cohen$^{\footnotesize{1,2}}$, Andreas Menzel$^{\footnotesize{2,3}}$
and Gal deBotton$^{\footnotesize{1,4}}$\\
$^{\footnotesize{1}}${\footnotesize{Dept.~of Mechanical Engineering, Ben-Gurion University, Beer-Sheva 84105, Israel;}}\\[-8 mm]
$^{\footnotesize{2}}${\footnotesize{Institute of Mechanics,~Dept. of Mechanical Engineering, TU Dortmund, D-44227 Dortmund, Germany;}}\\[-8 mm]
$^{\footnotesize{3}}${\footnotesize{Division of Solid Mechanics, Lund University, P.O.Box 118, SE-22100 Lund, Sweden;}}\\[-8 mm]
$^{\footnotesize{4}}${\footnotesize {Dept.~of Biomedical Engineering, Ben-Gurion University, Beer-Sheva 84105, Israel;}}}
\end{onehalfspace}
\maketitle
\begin{abstract}
\begin{onehalfspace}
Due to the increasing number of industrial applications of electro-active
polymers (EAPs), there is a growing need for electromechanical models
which accurately capture their behavior. To this end, we compare the
predicted behavior of EAPs undergoing homogenous deformations according
to three electromechanical models. The first model is a continuum
based model composed of the mechanical Gent model and a linear relationship
between the electric field and the polarization. The electrical and
the mechanical responses according to the second model are based on
the polymer microstructure, whereas the third model incorporates a
neo-Hookean mechanical response and a microstructural based long-chains
model for the electrical behavior. In the microstructural motivated
models the integration from the microscopic to the macroscopic levels
is accomplished by the micro-sphere technique. Four types of homogenous
boundary conditions are considered and the behaviors determined according
to the three models are compared. The differences between the predictions
of the models are discussed, highlighting the need for an in-depth
investigation of the relations between the structure and the behaviors
of the EAPs at microscopic level and their overall macroscopic response. \end{onehalfspace}

\end{abstract}
Keywords: dielectrics, EAPs, electromechanical coupling, multi-scale
analysis

\footnotetext{Author for correspondence: N. Cohen, e-mail: noyco@post.bgu.ac.il}

\begin{onehalfspace}

\section{Introduction}
\end{onehalfspace}

\begin{onehalfspace}
\vspace{7 mm}Dielectric elastomers (DEs) are materials that deform
under electrostatic excitation. Due to their light weight, flexibility
and availability these materials can be used in a wide variety of
applications such as artificial muscles \citep{barc01book}, energy-harvesting
devices \citep{mcka&etal10apl,Springhetti01102014}, micropumps \citep{rudy&etal12ijnm},
and tunable wave guides \citep{0964-1726-22-10-104014,Shmuel2012307},
among others. The principle of the actuation is based on the attraction
between two oppositely charged electrodes attached to the faces of
a thin soft elastomer sheet. Due to Poisson's effect, the sheet expands
in the transverse direction. \citet{toup56arma}, in his theoretical
work, found that this electromechanical coupling is characterized
by a quadratic dependence on the applied electric field. This was
later verified experimentally by \citet{kofo&etal03jimss}. However,
DEs have a low energy-density in comparison with other actuators such
as piezoelectrics and shape memory alloys \citep{barc01book}. Furthermore,
their feasibility is limited due to the high electric fields ($\sim100\,\mathrm{MV/m}$)
required for a meaningful actuation as a result of the relatively
low ratio between the dielectric and elastic moduli \citep{Pelrine200089,Pelrine04022000}.
Specifically, common flexible polymers have low dielectric moduli
while polymers with high dielectric moduli are usually stiff. Nevertheless,
a few recent works suggest that this ratio may be improved. \citet{huan&etal04aple}
demonstrated experimentally that organic composite EAPs (electro-active
polymers) experience more than 8\% actuation strain in response to
an activation field of $20\,\mathrm{MV/m}$. The experimental work
of \citet{stoy&etal10jmatchem} showed that the actuation can be dramatically
improved by embedding conducting particles in a soft polymer. In parallel,
theoretical works dealing with the enhancement of coupling in composites
also hint at the possibility of improved actuation with an appropriate
adjustment of their microstructure \citep{tian12jmps,Galipeau20121,:/content/aip/journal/apl/102/15/10.1063/1.4801775,0964-1726-22-10-104014,lopez2014elastic}. 

The above findings motivate an in-depth multiscale analysis of the
electromechanical coupling in elastic dielectrics which is inherent
from their microstructure. In this work we consider the class of polymer
dielectrics. A polymer is a hierarchical structure of polymer chains
each of which is a long string of repeating monomers. We start by
analyzing the behavior of a single monomer in a chain. Next, the response
of a chain is obtained by a first level integration from the single-monomer
level to the chain level. Finally, the macroscopic behavior of the
polymer is obtained by a higher level summation over all chains. In
this work, we utilize existing constitutive models for the chains
and concentrate on the higher level summation from the chain level
to the macroscopic-continuum level. To this end the physically motivated
micro-sphere technique, that enables to extend one dimensional models
to three dimensional models by appropriate integration over the orientation
space, is exploited \citep{bavzant1986efficient,Carol2004511}. Accordingly,
this method lends itself to the characterization of polymer networks
since the single chain is often treated as a 1-D object which is aligned
along the chain's end-to-end vector \citep{Miehe20042617,0964-1726-21-9-094008}.

The response of a polymer subjected to purely mechanical loadings
was extensively investigated at all length scales. A \emph{macroscopic
level }analysis and models describing the behavior of soft materials
undergoing large deformation, such as polymers, were developed by
\citet{ogden97book}. The \emph{microscopic level }analysis of \citet{kuhn1942beziehungen}
yielded a Langevin based constitutive relation and paved the way to
various multiscale models such as the 3-chain model \citep{wang&guth52jcp},
the tetrahedral model \citep{:/content/aip/journal/jcp/11/11/10.1063/1.1723791,TF9464200083}
and the 8-chain model \citep{arru&boyc93jmps}. Corresponding micro-sphere
implementation of the Langevin model was carried out by \citet{Miehe20042617}. 

The electric response of dielectrics to electrostatic excitation was
examined by \citet{tier90book} and \citet{hutt&etal06book}, among
others, macroscopically as well as through their microstructure. Starting
with the examination of a single charge under an electric field, the
relations between the different electric macroscopic quantities, such
as the electric displacement, the electric field and the polarization,
and microscopic quantities, such as the free and bound charge densities
and the dipoles were defined and analyzed.

The study of the response of dielectrics to a coupled electromechanical
loading initiated with the pioneering work of \citet{toup56arma},
who performed a theoretical analysis at the \emph{macroscopic level.
}Later on, an invariant-based representation for the constitutive
behavior of EAPs was introduced by \citet{dorf&ogde05acmc}. Subsequently,
\citet{Ask2012156,Ask20129}, \citet{0964-1726-22-10-104014} and
\citet{jimenez2013deformation} investigated the possible influence
of the deformation and its rate on the electromechanical coupling.
\citet{0964-1726-21-9-094008} made use of a corresponding micro-sphere
technique at the chain level. By employing macroscopic constitutive
models for the mechanical and the electrical behavior of the polymer
chains, a few boundary value problems were solved by means of the
numerical implementation of the micro-sphere technique and the finite
element method. Initial multiscale analyses of the electromechanical
response were performed by \citet{Cohen14b,Cohen2014}. The present
work focuses on the implementation of different electromechanical
models to EAPs experiencing homogenous deformations under various
types of boundary conditions and examination of their predicted response.

We begin this work with the detailed description of the different
macroscopic and microscopic models and the presentation of the micro-sphere
framework. Next, the micro-sphere technique is used to compute the
macroscopic behavior of dielectrics with randomly oriented and uniformly
distributed dipoles experiencing the macroscopic rotation. In section
4 we determine and compare the behavior of a polymer according to
three electromechanical models under different boundary conditions.
The conclusions are gathered in section 5. \vspace{7 mm}
\end{onehalfspace}

\begin{onehalfspace}

\section{Theoretical background}
\end{onehalfspace}

\begin{onehalfspace}
\vspace{7 mm} 
Consider the deformation of a hyperelastic dielectric body subjected to electro-mechanical loading from a referential configuration to a current one.
In the reference configuration, the body occupies a region $V_{0}\in{\mathbb{R}^{3}}$ with a boundary $\partial V_{0}$.  The referential location of a material point is $\refposT$.
In the current configuration, the body occupies a region $V\in{\mathbb{R}^{3}}$  with a boundary $\partial V$, and we denote the location of a material point by $\curposT$.
The mapping of positions of material points from the reference to the current configurations is  $\curposT=\mapping(\refposT)$, and the corresponding deformation gradient is
\begineq{deformation gradient}
\defgT = \nabla_{\refposT}\mapping, 
\fineq
where the operation $\nabla_\refposT$ denotes the gradient with respect to the referential coordinate system. 
The right and left Cauchy-Green strain tensors are 
$\CGstrainT = \defgT ^{T} \defgT$ 
and 
$\CGleftstrainT = \defgT \, \defgT^{T}$. 
The ratio between the volumes of an infinitesimal element in the current and the reference configurations, $J = \det{\defgT}$,  is strictly positive. In the case of incompressible materials, which are of interest in the present work, we have $J=1$.

The induced electric field $\EfieldT$ on the body satisfies the governing equation 
\begineq{curl free E}
\nabla_{\curposT}\times\EfieldT=\mathbf{0}, 
\fineq
in the entire space, where $\nabla_\curposT$ denotes the gradient with respect to the current coordinate system.
Consequently, we can define a scalar field, the electric potential $\electricpotential$, such that $\EfieldT= -\nabla_{\curposT} \electricpotential$. 
The electric displacement field is 
\begineq{dielectric displacement}
\DfieldT = \vacpermittivity\EfieldT + \polarizationT,
\fineq
where  $\vacpermittivity$ is the permittivity of the vacuum and $\polarizationT$ is the polarization, or the electric dipole-density.
We recall that in vacuum $\polarizationT = \mathbf{0}$.
In the absence of free charges the electric displacement field is governed by the local equation
\begineq{divergence free d} 
\nabla_{\curposT}\cdot\DfieldT=0.
\fineq 
In the work of \cite{dorf&ogde05acmc}, the referential counterparts $\EfieldTref$ and $\DfieldTref$ of the electric field and the electric displacement were determined.  Specifically, 
\begineq{referential electric field}  
\EfieldTref=\defgT^{T}\EfieldT,
\fineq 
\begineq{referential electric displacement} 
\DfieldTref=J\defgT^{-1}\DfieldT, 
\fineq 
where  $\nabla_\refposT\times\EfieldTref=\mathbf{0}$  and $\nabla_\refposT\cdot\DfieldTref=0$.   
We note that unlike $\EfieldTref$ and $\DfieldTref$,  the referential polarization is not uniquely defined. 
In order to ensure that the referential polarization is energy conjugate to the referential electric field such that $\frac{1}{J}\EfieldTref\cdot\polarizationTref=\EfieldT\cdot\polarizationT$, we adapt the definition
\begineq{referential polarization} 
\polarizationTref=J\defgT^{-1}\polarizationT. 
\fineq 

In accordance with our assumption that the dielectric solid can be treated as a hyper-elastic material, its constitutive behavior can be characterized in terms of a scalar electrical enthalpy per unit volume function $\energy$.
We further assume that $\energy$ can be decomposed into a mechanical and a coupled contributions, i.e. 
$\energy\left(\defgT,\EfieldT\right)=
\energy_{0}\left(\defgT\right)+\energy_{c}\left(\defgT,\EfieldT\right)$, 
where $\energy_{0}\left(\defgT\right)$ characterizes the material response in the absence of electric excitation and $\energy_{c} \left(\defgT,\EfieldT\right)$ accounts for the difference between $\energy$ with and without electric excitation \citep{mcme&land05jamt, mcme&etal07ijnm, Cohen2014, Cohen14b}.
Accordingly, the polarization is determined via
\begineq{polarization definition}
\polarizationT = -\frac{1}{J}  \frac{\partial \energy_{c}} {\partial \EfieldT},
\fineq
and the stress developing in the material can be written as the sum
\begineq{stress decomposition}
\stressT = \mechanicalstressT + \electricstressT + \maxwellstressT,
\fineq
where
\begineq{mechanical stress}
\mechanicalstressT=\frac{1}{J}\, \frac{\partial{\energy\left(\defgT\right)}}{\partial{\defgT}}\,\defgT^{T},
\fineq
is the mechanical stress due to the deformation of the material,
\begineq{polarization stress}
\electricstressT = \EfieldT \otimes \polarizationT,
\fineq
is the polarization stress stemming from the applied electric field in the dielectric, and
\begineq{maxwell stress}
\maxwellstressT = \vacpermittivity \left[\EfieldT\otimes\EfieldT - 
\frac{1}{2} \left[\EfieldT\cdot\EfieldT \right] \mathbf{I} \right],
\fineq
is the Maxwell stress in vacuum, where $\mathbf{I}$  is the second order identity tensor \citep{mcme&land05jamt}.
We emphasize that this decomposition is purely modelling-based as in an experiment  the total stress can be measured, but the contributions of the individual components cannot be distinguished.
In this work, we consider incompressible materials that undergo homogenous deformations and therefore a pressure like term $p\, \mathbf{I}$, which is determined from the boundary conditions, is added to the total stress.
Assuming no body forces, the stress satisfies the local equilibrium equation 
\begineq{equilibrium} 
\nabla_{\curposT}\cdot\stressT=\mathbf{0}.
\fineq  

The electrical boundary conditions are given in terms of either the electric potential or the charge per unit area $\rho_a$, such that $\DfieldT\cdot\hat{\mathbf{n}} = -\rho_a$,
where $\hat{\mathbf{n}}$ is the outward pointing unit normal.   
Practically, in EAPs, $\rho_a$ is the charge on the electrodes.   
The mechanical boundary conditions are given in terms of the displacement or the mechanical traction $\tractionT$. 
Due to the presence of the electric field the stress in the vacuum outside of the body does not vanish.
Therefore, the mechanical traction at the boundary is
$\left[\stressT-\maxwellstressT \right]\cdot \hat{\mathbf{n}}=\tractionT$, where the expression for $\maxwellstressT$, the Maxwell stress outside the material, is given in \Eq{maxwell stress} in terms of the electric field in the vacuum.
\vspace{7 mm}
\end{onehalfspace}

\begin{onehalfspace}

\subsection{Existing models for the behavior of dielectrics}
\end{onehalfspace}

\begin{onehalfspace}
\vspace{7 mm}Within the framework of finite deformation elasticity
the simplest constitutive model is the well-known neo-Hookean model
that requires only one material parameter. The corresponding strain
energy-density function (SEDF) is 
\begin{equation}
\energy_{0}^{\,\, nH}\left(\defgT\right)=\frac{\shear}{2}\,\left[I_{1}-3\right],\label{eq:neo-Hookean}
\end{equation}
where $\shear$ is the shear modulus and $I_{1}=\mathrm{tr}\left(\CGstrainT\right)$
is the first invariant of the right Cauchy-Green strain tensor. This
model does not capture the \emph{lock-up} effect observed in experiments
and corresponds to a significant stiffening of the material at large
strains. \citet{gent96rc&t} proposed a phenomenological constitutive
model in which this effect is accounted for. The SEDF for this model
\begin{equation}
\energy_{0}^{\,\, G}\left(\defgT\right)=-\frac{\shear J_{m}}{2}\ln\left(1-\frac{I_{1}-3}{J_{m}}\right),\label{eq:gent}
\end{equation}
depends on two parameters, $\shear$ and $J_{m}$. The latter is the
lock-up parameter such that $J_{m}+3$ is the value of $I_{1}$ at
the lock-up stretch. Thus, the expression in Eq. (\ref{eq:gent})
becomes unbounded at $I_{1}=J_{m}+3$ which captures this phenomenon. 

With regard to the response of the dielectric to electrical excitation
a quadratic dependence of $\energy_{c}$ on $\EfieldT$ is commonly
assumed, leading to the linear relation 
\begin{equation}
\polarizationT=\susceptibility\EfieldT,
\end{equation}
where $\susceptibility$ is the susceptibility of the material \citep{blyt&bloo08book}.
Consequently, the electric displacement is 
\begin{equation}
\DfieldT=\permittivity\EfieldT,\label{eq:linear_relation_displacement}
\end{equation}
where $\permittivity=\vacpermittivity+\susceptibility$ is the permittivity
of the material. We note that this linear relation is in agreement
with the invariant based representation of \citet{dorf&ogde05acmc}.
Furthermore, an experiment carried out by \citet{:/content/aip/journal/jap/111/2/10.1063/1.3676201}
on VHB 4910 showed that this assumption is fairly accurate.

Recent experiments with various types of polymers imply that the permittivity,
and therefore the relation between the polarization and the electric
field, is deformation dependent \citep{choi2005effects,wissler2007electromechanical,mcka,qiang2012experimental}.
A possible explanation for this dependency of the susceptibility on
the deformation is related to the alteration of the inner structure
of the polymer \citep{Cohen2014,Cohen14b}. 

In the polymer, the monomers in the chain can move or rotate relative
to their neighbors thus providing the chains with a freedom to deform
\citep{flor53book}. In order to better understand the response of
polymers to an electro-mechanical loading, their microstructure should
be accounted for. This can be accomplished in terms of a multiscale
analysis consisting of three stages: the first involves the examination
of the behavior of the monomers, the second includes analysis of the
response of the chains, and the third deals with the polymer behavior
at the continuum level.

\citet{trel43atfaradaysoc} and \citet{:/content/aip/journal/jcp/11/11/10.1063/1.1723791}
carried out a multiscale analysis of a polymer subjected to mechanical
loading. It was assumed that the directions of the monomers, or links,
composing a chain are random, and consequently it was found that the
chains are distributed according to a Gaussian distribution. Based
on statistical considerations and the laws of thermodynamics, the
variation in the entropy of the chain due to its deformation was determined.
The overall variation in the entropy of the polymer is computed by
summing the entropies of the chains. Remarkably, their result recovered
the macroscopic neo-Hookean behavior. Furthermore, a comparison between
the micro and macro analyses related the macroscopic shear modulus
to the number of chains per unit volume $\chainspervolume_{0}$. Specifically,
it was found that 
\begin{equation}
\shear=k\, T\,\chainspervolume_{0},\label{eq:shear_modulus_micro}
\end{equation}
where $k$ and $T$ are the Boltzmann constant and the absolute temperature,
respectively.

Due to the assumptions that lead to the use of Gaussian statistics,
the\emph{ }lock-up\emph{ }effect was not captured in the above mentioned
analysis. A more rigorous examination of the polymer behavior by \citet{kuhn1942beziehungen}
revealed that this phenomenon is a result of the finite extensibility
of the chains. According to this analysis the SEDF associated with
a polymer chain is 
\begin{equation}
\energy_{_{0}}^{\,\, LC}=k\, T\,\numberoflinks\left[\invlangevin\left(\frac{\chainlencur}{\numberoflinks\,\linklen}\right)\,\frac{\chainlencur}{\numberoflinks\,\linklen}+\ln\left(\frac{\invlangevin\left(\frac{\chainlencur}{\numberoflinks\,\linklen}\right)}{\sinh\left(\invlangevin\left(\frac{\chainlencur}{\numberoflinks\,\linklen}\right)\right)}\right)\right],\label{eq:first langevin}
\end{equation}
where $\linklen$ is the length of a link, $\numberoflinks$ is the
number of links in a chain, $\chainlencur$ is the distance between
the two ends of the chain, and $\invlangevin\left(\bullet\right)$
is the inverse of the Langevin function 
\begin{equation}
\langevin\left(\beta\right)\equiv\coth\left(\beta\right)-\frac{1}{\beta}=\frac{\chainlencur}{\numberoflinks\,\linklen}.
\end{equation}
Assuming that all chains undergo the macroscopic deformation, i.e.
$\chainveccur=\defgT\,\chainvecref$ where $\chainveccur$ and $\chainvecref$
are the current and referential end-to-end vectors, respectively,
the stress associated with a chain is derived from Eq. (\ref{eq:first langevin}),
\begin{equation}
\mechanicalstressT^{\,\, LC}=k\, T\,\sqrt{\numberoflinks}\:\frac{\chainlenref}{\chainlencur}\,\invlangevin\left(\frac{\chainlencur}{\numberoflinks\,\linklen}\right)\,\defgT\hat{\chainveccur}_{0}\otimes\defgT\hat{\chainveccur}_{0},\label{eq:Langevin stress}
\end{equation}
where $\hat{\chainveccur}_{0}$ is a unit vector in the direction
of $\chainvecref$ and Eq. (\ref{mechanical stress}) is used. Here,
\begin{equation}
\chainlenref=\linklen\,\sqrt{\numberoflinks},\label{eq:referential_chain_len}
\end{equation}
is the average length of the referential end-to-end vectors \citep{flor53book,trel75book}.
The quantity $\frac{\chainlencur}{\numberoflinks\,\linklen}\leq1$
describes the ratio between the end-to-end and the contour lengths
of the chain, and from Eq. (\ref{eq:referential_chain_len}) it follows
that 
\begin{equation}
\frac{\chainlencur}{\numberoflinks\,\linklen}=\sqrt{\defgT\hat{\chainveccur}_{0}\cdot\defgT\hat{\chainveccur}_{0}}\,\frac{\chainlenref}{\numberoflinks\,\linklen}=\sqrt{\defgT\hat{\chainveccur}_{0}\cdot\defgT\hat{\chainveccur}_{0}}\,\frac{1}{\sqrt{\numberoflinks}}.
\end{equation}
It can be shown that the limit $\frac{\chainlencur}{\numberoflinks\,\linklen}\rightarrow1$
results in $\invlangevin\left(\frac{\chainlencur}{\numberoflinks\,\linklen}\right)\rightarrow\infty$,
thus capturing the experimentally observed lock-up phenomenon. The
lock-up stretch is associated with the chain undergoing the largest
extension such that the stretch ratio of its end-to-end vector is
\begin{equation}
\stretch_{max}=\sqrt{\numberoflinks}.\label{eq:Langevin_lock_up}
\end{equation}
It is important to note that the first term in the Taylor series expansion
of Eq. (\ref{eq:Langevin stress}) about $\frac{\chainlencur}{\numberoflinks\,\linklen}=0$
reproduces the Gaussian model \citep{kuhn1942beziehungen,trel75book}.

A few works proposed models that consider specific finite networks
of chains. The 3-chain model by \citet{wang&guth52jcp} examines a
network of 3 chains which are located along the axis of the principal
directions of the deformation gradient. \citet{:/content/aip/journal/jcp/11/11/10.1063/1.1723791}
and \citet{TF9464200083} proposed a network of four chains that are
linked together at the center of a regular tetrahedron, and their
other ends are located at the vertices of the tetrahedron. The tetrahedron
deforms according to the macroscopic deformation while the chains
experience different stretches. In the model proposed by \citet{arru&boyc93jmps},
8 representative chains in specific directions relative to the principal
system of the macroscopic deformation gradient are used to determine
the macroscopic behavior. An anisotropic worm-like chain model in
which no inherent alignment between the chosen and the principal coordinate
systems is assumed was considered by \citet{doi:10.1080/14786430500080296}.

A multiscale level analysis of the response of polymers with Gaussian
distribution to electro-mechanical loading was carried out by \citet{Cohen14b}.
In this study, the changes in the magnitudes of the dipolar monomers
due to the applied electric field and their rearrangement due to the
mechanical deformation were accounted for. Following a model described
in \citet{stockmayer1967dielectric}, \citet{Cohen2014} considered
the class of uniaxial dipoles in which the dipole is aligned with
the line segment between the two contact points of a monomer to its
neighbors. Thus, taking $\dipoledir$ to be the unit vector along
this line segment the dipole moment of a monomer is 
\begin{equation}
\dipoleT_{u}=\constant\left[\dipoledir\otimes\dipoledir\right]\EfieldT,\label{eq:uniaxial_dipole}
\end{equation}
where $\constant$ is a material constant. The dipole of a chain composed
of $\numberofdipoles$ uniaxial dipoles with an end-to-end vector
in the direction $\hat{\chainveccur}$ is \citep{Cohen14b} 
\begin{equation}
\dipoleT_{c}\approx\frac{\constant\,\numberofdipoles}{3}\left[\mathbf{I}+\frac{16}{3\,\pi^{2}}\left[\frac{\chainlencur}{\numberofdipoles\,\linklen}\right]\left[\mathbf{I}-3\hat{\chainveccur}\otimes\hat{\chainveccur}\right]\right]\EfieldT.\label{eq:long_chains_stress}
\end{equation}
If we assume that all chains undergo the macroscopic deformation,
then $\hat{\chainveccur}=\frac{\defgT\,\chainvecref}{\sqrt{\left[\defgT\,\chainvecref\right]\cdot\left[\defgT\,\chainvecref\right]}}$
. In accordance with a second type of dipoles discussed in \citet{stockmayer1967dielectric},
\citet{Cohen2014} proposed an expression for a transversely isotropic
(TI) dipole, where the dipole is aligned with the projection of the
electric field on the plane perpendicular to $\dipoledir$, 
\begin{equation}
\dipoleT_{t}=\frac{\constant}{2}\left[\mathbf{I}-\dipoledir\otimes\dipoledir\right]\EfieldT.\label{eq:TI dipoles}
\end{equation}
An expression for the dipole of a chain made out of transversely isotropic
dipoles is determined by following the steps followed in the derivation
of Eq. (\ref{eq:long_chains_stress}) for the uniaxial dipoles.

The resulting polarization of the polymer is determined by summing
the dipoles of the chain in a representative volume element of a volume
$V^{R}$ via \citep{blyt&bloo08book}
\begin{equation}
\polarizationT=\frac{1}{V^{R}}\sum_{i}\dipoleT_{c},\label{eq:polarization_definition}
\end{equation}
and the resulting polarization stress is computed via Eq. (\ref{polarization stress}).\vspace{7 mm}
\end{onehalfspace}

\begin{onehalfspace}

\subsection{The micro-sphere technique}
\end{onehalfspace}

\begin{onehalfspace}
\vspace{7 mm}Consider a unit sphere whose surface represents the
directions of the referential end-to-end vectors. The directional
averaging of a quantity $\bullet$ over the unit sphere can be approximated
with a discrete summation 
\begin{equation}
\left\langle \bullet\right\rangle =\frac{1}{4\pi}\intop_{A}\,\bullet\,\d A\approx\sum_{i=1}^{m}\bullet^{\left(i\right)}w^{\left(i\right)},\label{eq:micro_sphere_averaging}
\end{equation}
where the index $i=1,...,m$ refers to a unit direction vector $\indexunit$
where $\bullet^{\left(i\right)}$ is the value of quantity $\bullet$
in the direction $\indexunit$ and $w^{\left(i\right)}$ is an appropriate
non-negative weight function constrained by $\Sigma_{i=1}^{m}w_{i}=1$
\citep{bavzant1986efficient,Miehe20042617,menzel2009microsphere}.
In general, Eq. (\ref{eq:micro_sphere_averaging}) can be combined
with a more general anisotropic distribution function \citep{Alastru2009178,NME:NME2577}.

In the case of polymers the vectors $\indexunit$ represent the directions
of the end-to-end vectors of the polymer chains or, from a numerical
point of view, the integration directions in orientation space. For
randomly and isotropically oriented chains, or rather isotropic integration
schemes, the vectors $\indexunit$ satisfy 
\begin{equation}
\sum_{i=1}^{m}\,\indexunit\, w^{\left(i\right)}=\mathbf{0},\label{eq:random 1}
\end{equation}
and 
\begin{equation}
\sum_{i=1}^{m}\,\indexunit\otimes\indexunit\, w^{\left(i\right)}=\frac{1}{3}\,\mathbf{I}.\label{eq:random vectors tensor multiplication}
\end{equation}
In view of Eqs. (\ref{eq:random 1}) and (\ref{eq:random vectors tensor multiplication})
the micro-sphere technique naturally lends itself to the calculation
of the macroscopic polarization and stress. Specifically, for a polymer
with chains composed of $\numberofdipoles$ dipoles and $N_{0}$ chains
per unit referential volume, Eq. (\ref{eq:polarization_definition})
may be written as

\begin{equation}
\polarizationT=\frac{N_{0}}{J}\left\langle \dipoleT_{c}\right\rangle ,\label{eq:polarization_approximation}
\end{equation}
and the macroscopic stress according to Eq. (\ref{eq:Langevin stress})
as
\begin{equation}
\mechanicalstressT^{\,\, L}=\frac{N_{0}}{J}\left\langle \mechanicalstressT^{\,\, LC}\right\rangle ,\label{eq:stress_approximation}
\end{equation}
where we use the notation suggested in Eq. (\ref{eq:micro_sphere_averaging}).

\citet{bavzant1986efficient} demonstrated that a specific choice
of $42$ directions guarantees sufficient accuracy for the application
discussed in their work. We follow this conjecture where the integration
directions and the corresponding weight functions are given in Table
1 of \citet{bavzant1986efficient}. We note that other integration
schemes are available, as demonstrated by \citet{Waffenschmidt20121928,NME:NME4601}
and the references cited therein.\vspace{7 mm}
\end{onehalfspace}

\begin{onehalfspace}

\section{Dielectrics with randomly distributed monomers}
\end{onehalfspace}

\begin{onehalfspace}
\vspace{7 mm}Consider a model of a dielectric composed of $n_{0}$
monomers per unit referential volume, which are treated as mechanical
rods and electric dipoles. The dielectric is subjected to a mechanical
deformation, locally represented by $\defgT$, and an electric field
$\EfieldT$. We assume that the electric field induced on a monomer
by its neighbors is small in comparison with the applied electric
field \citep{Cohen2014}. We examine first a dielectric with uniaxial
monomers, the behavior of which is governed by the quadratic form
in Eq. (\ref{eq:uniaxial_dipole}). If all of the dipoles experience
the macroscopic rotation, i.e. $\dipoledir=\rotation\,\dipoledirref$
where $\rotation=\defgT\,\CGstrainT^{-1/2}$ is a proper rotational
tensor, then the polarization according to Eq. (\ref{eq:polarization_approximation})
is 
\begin{equation}
\polarizationT=n_{0}\left\langle \dipoleT_{c}\right\rangle =n_{0}\,\constant\,\rotation\,\sum_{i=1}^{42}\dipoledirref^{\left(i\right)}\otimes\dipoledirref^{\left(i\right)}\, w^{\left(i\right)}\,\rotation^{T}\,\EfieldT=\frac{n_{0}\,\constant}{3}\,\EfieldT,\label{eq:Puniaxial}
\end{equation}
where Eqs. (\ref{eq:micro_sphere_averaging}) and (\ref{eq:random vectors tensor multiplication})
are used. The corresponding polarization stress is 
\begin{equation}
\electricstressT=\EfieldT\otimes\polarizationT=\frac{n_{0}\,\constant}{3}\,\EfieldT\otimes\EfieldT.\label{eq:stress_dielectric}
\end{equation}

In the case of a dielectric composed of $n_{0}$ TI dipolar monomers
per unit referential volume, cf. Eq. (\ref{eq:TI dipoles}), which
mechanically act as rigid rods, the same assumptions that led to Eq.
(\ref{eq:Puniaxial}) lead to
\begin{equation}
\polarizationT=n_{0}\left\langle \dipoleT_{c}\right\rangle =n_{0}\,\frac{\constant}{2}\sum_{i=1}^{42}\left[\mathbf{I}-\rotation\,\dipoledirref^{\left(i\right)}\otimes\dipoledirref^{\left(i\right)}\rotation^{T}\right]w^{\left(i\right)}\EfieldT=\frac{n_{0}\,\constant}{3}\,\EfieldT.\label{eq:PTI}
\end{equation}
Accordingly, the expression for the polarization stress is given in
Eq. (\ref{eq:stress_dielectric}). We note that the polarization and
polarization stress calculated in Eqs. (\ref{eq:Puniaxial}), (\ref{eq:PTI})
and (\ref{eq:stress_dielectric}) are identical to the exact expressions
obtained by \citet{Cohen2014}.\vspace{7 mm}
\end{onehalfspace}

\begin{onehalfspace}

\section{Dielectric elastomers}
\end{onehalfspace}

\begin{onehalfspace}
\vspace{7 mm}We examine the behaviors of incompressible dielectric
elastomers according to three different models under various homogenous
electromechanical loading conditions and compare between their predicted
responses. To facilitate the comparison we assume that in the limit
of infinitesimal deformations and small electric excitations all three
models admit the same behavior. Specifically we assume that the initial
shear modulus and electric susceptibility are $\shear=0.1\,\mathrm{MPa}$
and $\susceptibility=3\,\vacpermittivity$. In those models in which
the lock-up effect is accounted for we choose the model parameters
such that under purely mechanical biaxial loading the lock-up stretch
is $\stretch^{lu}=5$. The precise models and the numerical values
assumed for their parameters are as follows:
\end{onehalfspace}
\begin{enumerate}
\begin{onehalfspace}
\item The \textit{macroscopic} model - the mechanical behavior is characterized
by the Gent model (\ref{eq:gent}) with the aforementioned shear modulus
and $J_{m}=47$. The electric behavior is determined according to
the linear model (\ref{eq:linear_relation_displacement}) with the
initial permittivity $\permittivity=4\,\vacpermittivity$.
\item The \textit{microscopic} model - the Langevin model (\ref{eq:Langevin stress})
is utilized in order to describe the mechanical behavior, where $\numberoflinks=25$
is chosen to fit the assumed lock-up stretch according to Eq. (\ref{eq:Langevin_lock_up})
and $N_{0}=\frac{\shear}{k\, T}$. We employ the long-chains model
(\ref{eq:long_chains_stress}) with chains that are composed of $\numberofdipoles=100$
uniaxial dipoles (Eq. \ref{eq:uniaxial_dipole}) to characterize the
dielectric response of the polymer. The material constant $\constant$
is chosen such that $\frac{\constant\, N_{0}\,\numberofdipoles}{3}=3\,\vacpermittivity$
to ensure that the referential polarization is identical to the one
admitted by the macroscopic model. 
\item The \textit{Gaussian} model - the neo-Hookean model (\ref{eq:neo-Hookean})
with (\ref{eq:shear_modulus_micro}) are used, where $N_{0}=\frac{\shear}{k\, T}$,
in conjunction with the long-chains model (\ref{eq:long_chains_stress})
to characterize the mechanical and the electrical behaviors, respectively.
We assume that a chain is composed of $\numberofdipoles=100$ uniaxial
dipoles (Eq. \ref{eq:uniaxial_dipole}), where the chosen long-chains
model constant is identical to the one determined for the microscopic
model.\end{onehalfspace}

\end{enumerate}
\begin{onehalfspace}
In the following, we examine a thin layer of a polymer whose opposite
faces are covered with flexible electrodes with negligible stiffness.
The electrodes are charged with opposite charges so that the difference
in the electric potential induces an electric field across the layer.
From a mechanical point of view we consider four boundary conditions.
In the first two cases different displacements are prescribed at the
boundary and consequently the deformation gradient is defined. In
the following two representative cases we set the traction on the
boundaries. We choose a cartesian coordinate system in which the referential
electric field is aligned with the $\mathbf{\hat{y}}$-axis and calculate
the macroscopic polarization according to the microscopic and the
Gaussian models via Eq. (\ref{eq:polarization_approximation}). The
polarization stress is computed according to Eq. (\ref{polarization stress}).
The mechanical stress is computed via Eqs. (\ref{eq:gent}), (\ref{eq:neo-Hookean})
and (\ref{eq:Langevin stress}) for the macroscopic, the Gaussian
and the microscopic models, respectively. The pressure term is determined
from the traction free boundaries to which the electrodes are attached
and subsequently the total stress is computed via Eq. (\ref{stress decomposition}). 

\global\long\def\dimStress{\bar{\sigma}}
\global\long\def\dimEfield{\bar{\Efield}}
\global\long\def\dimDfield{\bar{D}}
\global\long\def\dimEref{\bar{\Efield}^{\left(0\right)}}
\global\long\def\dimDref{\bar{D}^{\left(0\right)}}

For convenience we define the dimensionless normal stress along the
$\mathbf{\hat{x}}$-axis $\dimStress=\frac{1}{\shear}\,\mathbf{\hat{x}}\cdot\stressT\,\mathbf{\hat{x}}$
and the dimensionless referential electric field and referential electric
displacement along the $\mathbf{\hat{y}}$-direction $\dimEref=\sqrt{\frac{\permittivity}{\shear}}\,\EfieldTref\cdot\hat{\mathbf{y}}$
and $\dimDref=\frac{1}{\sqrt{\permittivity\,\shear}}\,\DfieldTref\cdot\hat{\mathbf{y}}$,
respectively. In the following examples the current configuration
counterparts of $\dimEref$ and $\dimDref$ are $\dimEfield=\sqrt{\frac{\permittivity}{\shear}}\,\EfieldT\cdot\hat{\mathbf{y}}$
and $\dimDfield=\frac{1}{\sqrt{\permittivity\,\shear}}\,\DfieldT\cdot\hat{\mathbf{y}}$,
respectively, as follows from Eqs. (\ref{referential electric field})
and (\ref{referential electric displacement}).\vspace{7 mm}
\end{onehalfspace}

\begin{onehalfspace}

\subsection{Equibiaxial stretching perpendicular to the electric field\label{sub:Biaxial}}
\end{onehalfspace}

\begin{onehalfspace}
\vspace{7 mm}In this case the material is stretched along the axes
$\mathbf{\hat{x}}$ and $\mathbf{\mathbf{\hat{z}}}$ such that $\stretch_{x}=\stretch_{z}=\stretch$.
As stated previously, the $\mathbf{\hat{y}}$-axis is aligned with
the referential electric field and due to the assumed incompressibility
$\stretch_{y}=\frac{1}{\stretch^{2}}$. This setting is common in
various experiments with EAPs \citep{choi2005effects,wissler2007electromechanical,mcka,qiang2012experimental}.

\begin{figure}
\hspace*{\fill}\includegraphics[scale=0.45]{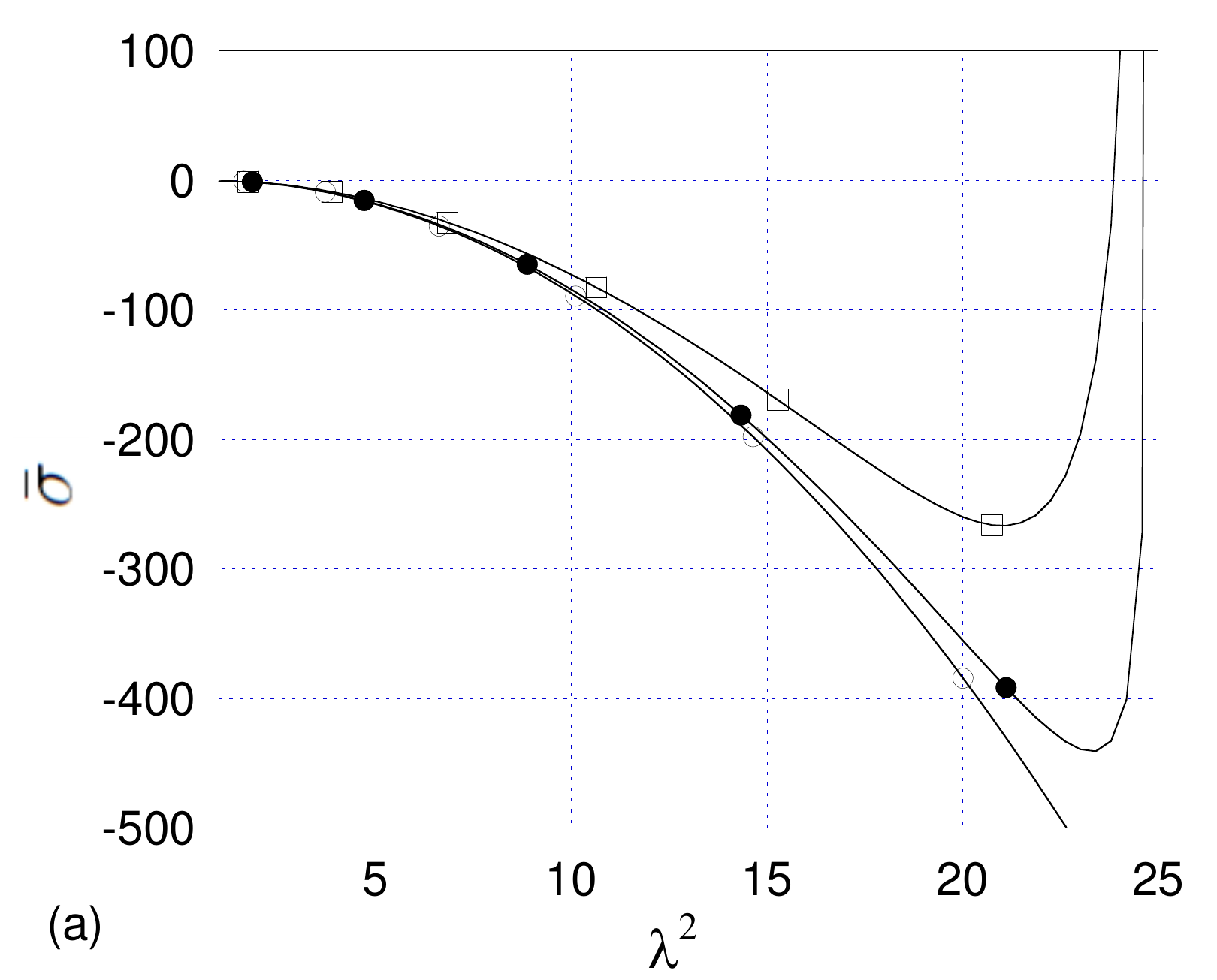}\qquad{}
\includegraphics[scale=0.45]{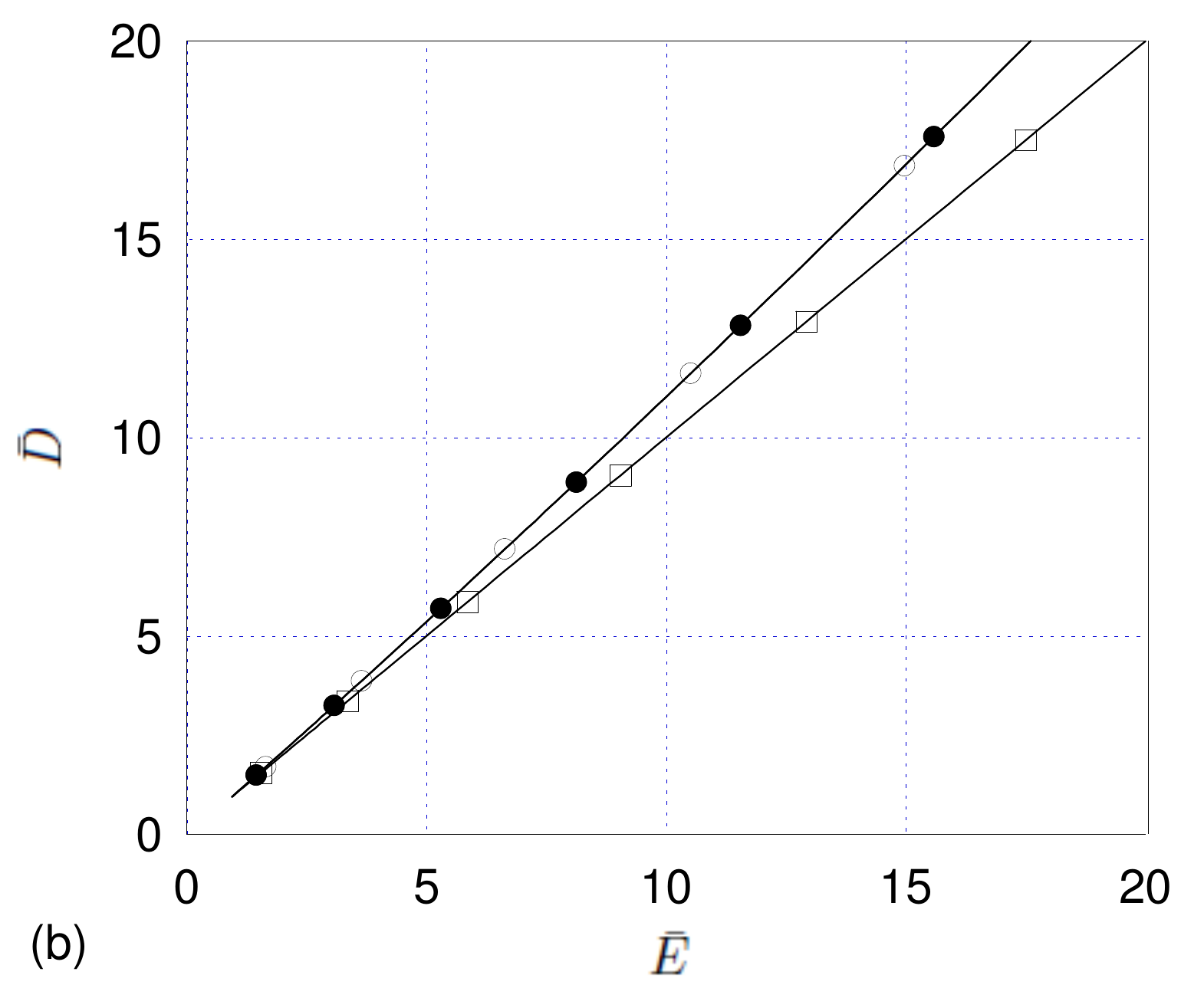}\hspace*{\fill}

\protect\caption{The dimensionless stress $\protect\dimStress$ (a) and electric displacement
$\protect\dimDfield$ (b) versus the stretch of the transverse plane
and the dimensionless electric field according to the macroscopic
model (the curve with the square marks), the Gaussian model (the curve
with the hollow circle marks) and the microscopic model (the curve
with the filled circle marks). \label{fig:biaxial_stretch}}
\end{figure}

Figs. (\ref{fig:biaxial_stretch}a) and (\ref{fig:biaxial_stretch}b)
depict $\dimStress$ and $\dimDfield$ as functions of $\stretch^{2}$
and $\dimEfield$, respectively. The curves with the squared marks
correspond to the macroscopic model, the curves with the hollow circle
marks to the Gaussian model, and the curves with the filled circle
marks to the microscopic model. The applied referential electric field
is $\EfieldRef=50\,\mathrm{\frac{MV}{m}}$. We point out that at $\stretch^{2}=1$
the dimensionless stress according to the three models is not zero
but very small. Fig. (\ref{fig:biaxial_stretch}a) illustrates the
stress increase at the lock-up stretch according to the macroscopic
and the microscopic models. As expected, this effect is not observed
when the Gaussian model is employed. Furthermore, since the electric
field tends to stretch the material in the transverse plane, as long
as the prescribed stretch is smaller than the electrically induced
stretch, the overall stress is compressive. We emphasize that since
the potential is held fixed, as the layer is stretched the current
electric field increases and hence also the electromechanically induced
stress. This gives rise to different types of loss of stability phenomena
which are outside the scope of the current work. The reader is referred
to the works by, e.g., \citet{Dorfmann20101,Bertoldi201118,:/content/aip/journal/apl/102/15/10.1063/1.4801775}
and \citet{Shmuel2012307}. Only when the prescribed stretches are
large enough, the total stress becomes tensile. 

In Fig. (\ref{fig:biaxial_stretch}b) we observe a linear dependence
of the electric displacement on the electric field according to the
macroscopic model, as follows from Eq. (\ref{eq:linear_relation_displacement})
and the assumed constant permittivity. Since the graph is plotted
in terms of the dimensionless quantities its slope is unity. In contrast,
the Gaussian and the microscopic models predict a stronger than linear
increase in the electric displacement as we stretch the material.
This is a result of the predicted increase in the permittivity due
to the stretching of a polymer with chains made up of uniaxial dipoles
\citep{Cohen14b}.

\vspace{7 mm}
\end{onehalfspace}

\begin{onehalfspace}

\subsection{Pure shear deformation in the plane of the electric field\label{sub:Pure shear}}
\end{onehalfspace}

\begin{onehalfspace}
\vspace{7 mm}Once again we analyze a thin layer of a polymer whose
opposite faces are covered with flexible electrodes with negligible
stiffness. As before the $\mathbf{\hat{y}}$-axis is aligned with
the electric field, but in this case the deformation of the material
along the $\mathbf{\mathbf{\hat{z}}}$-axis is constraint such that
$\stretch_{z}=1$. The layer is stretched along the $\mathbf{\hat{x}}$-axis
such that $\stretch_{x}=\stretch$ and the incompressibility condition
yields $\stretch_{y}=\frac{1}{\stretch}$.

\begin{figure}
\hspace*{\fill}\includegraphics[scale=0.45]{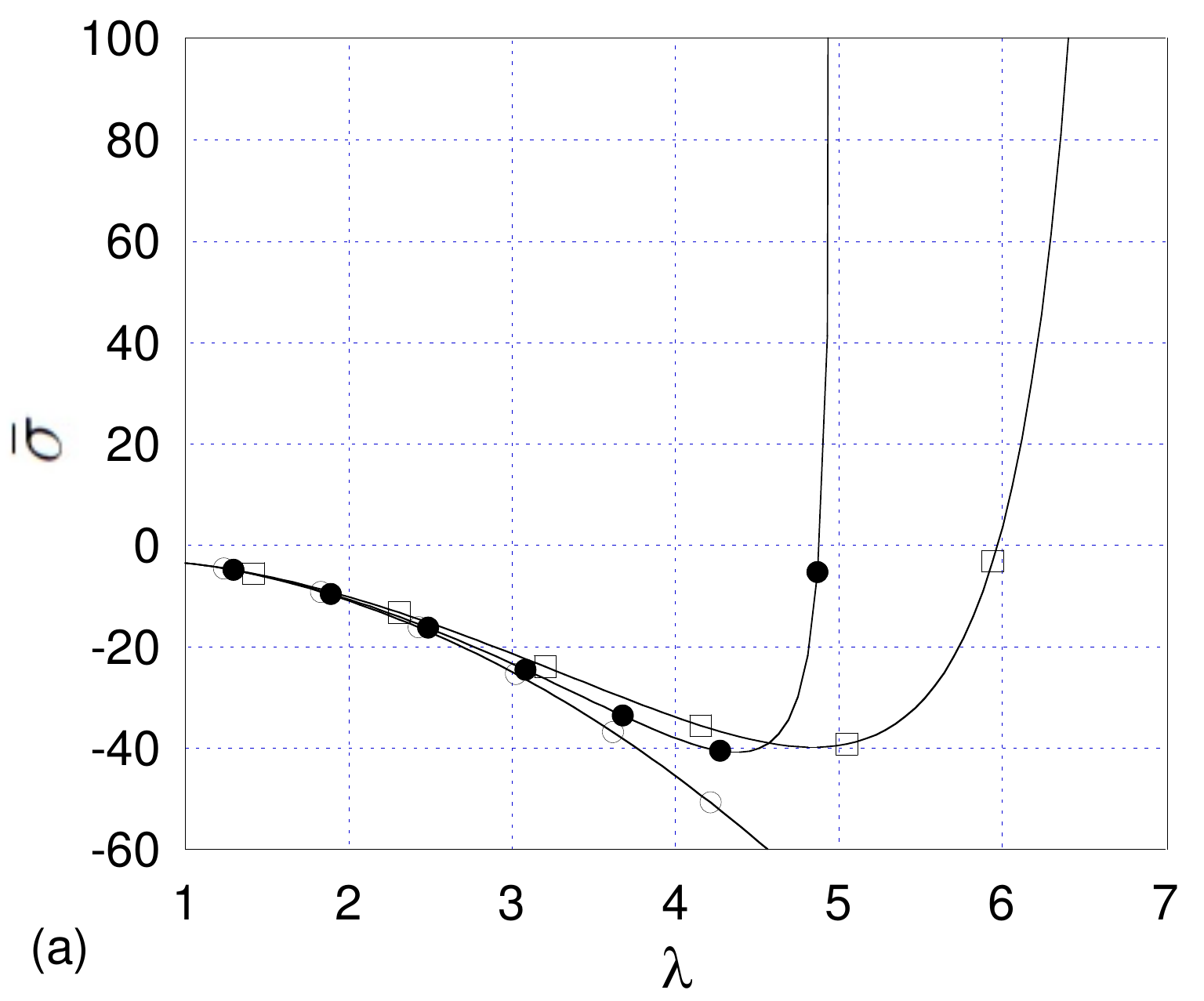}\qquad{}\includegraphics[scale=0.45]{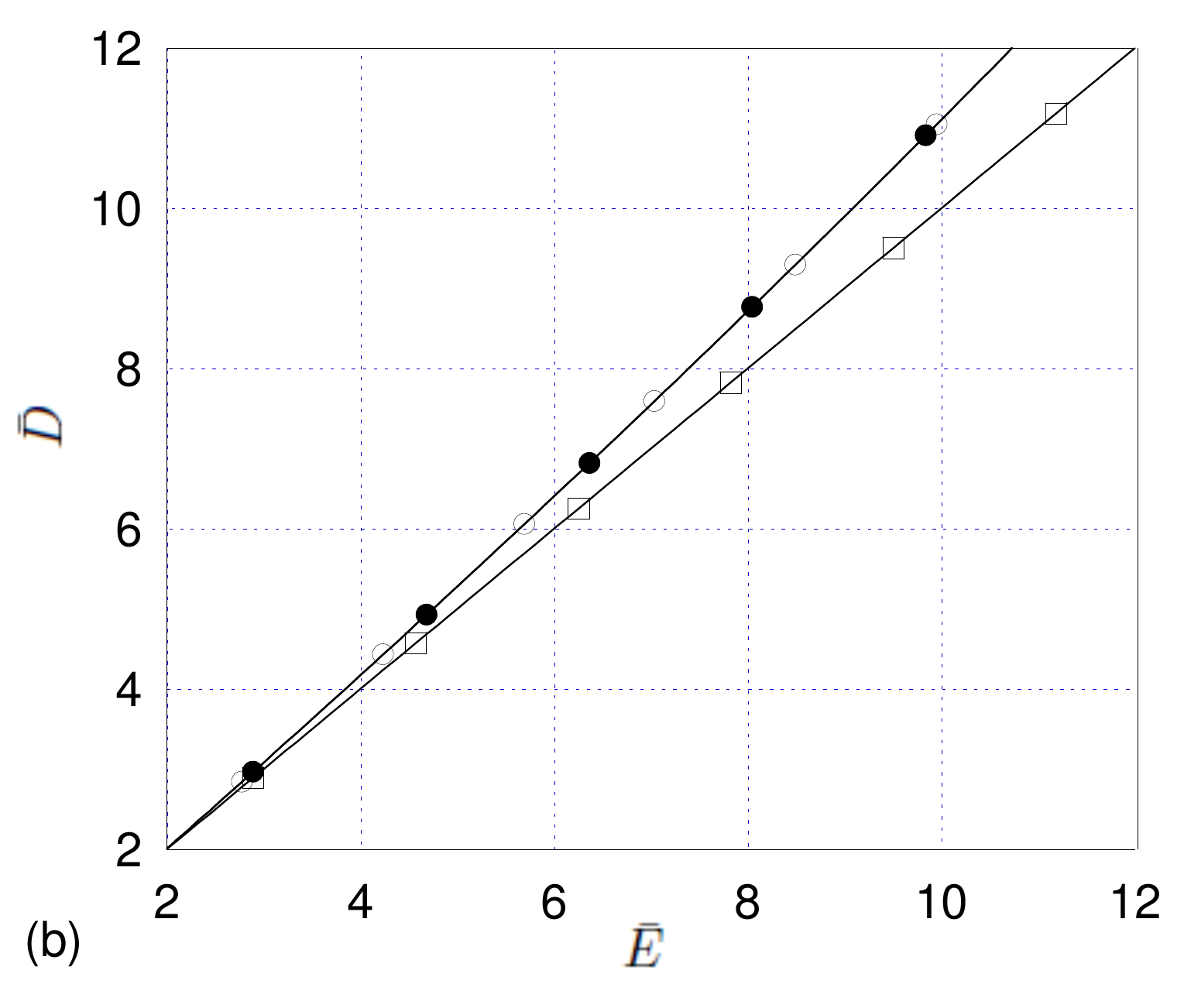}\hspace*{\fill}

\protect\caption{The dimensionless stress $\protect\dimStress$ (a) and electric displacement
$\protect\dimDfield$ (b) versus the axial stretch and the dimensionless
electric field according to the macroscopic model (the curve with
the square marks), the Gaussian model (the curve with the hollow circle
marks) and the microscopic model (the curve with the filled circle
marks). \label{fig:pure_shear}}
\end{figure}

Fig. (\ref{fig:pure_shear}a) depicts the dimensionless normal stress
that develops along the $\mathbf{\hat{x}}$-axis as a function of
the stretch according to the macroscopic model (the curve with the
square marks), the Gaussian model (the curve with the hollow circle
marks) and the microscopic model (the curve with the filled circle
marks) under the applied referential electric field $\EfieldRef=100\,\mathrm{\frac{MV}{m}}$.
The inability of the neo-Hookean model to capture the lock-up stretch
is again clearly depicted. We also notice that there is a difference
between the lock-up stretches predicted by the macroscopic and the
microscopic models. This is because the lock-up stretch according
to the Langevin model is determined by the maximum eigenvalue of the
deformation gradient as seen from Eq. (\ref{eq:Langevin_lock_up}),
whereas according to the Gent model it depends on the first invariant
of the right Cauchy-Green strain tensor. \citet{trel75book} presented
experimental results demonstrating that polymers lock-up at different
values under different types of deformations, and therefore we conclude
that in this aspect the Gent model may be a better predictor. We wish
to point out that the lock-up stretch according to the microscopically
motivated 8-chain model of \citet{arru&boyc93jmps}, in which the
chain behaves according to Eq. (\ref{eq:Langevin stress}), depends
on $I_{1}$ as well and is able to capture the different lock-up stretch
values under various states of deformation. A comparison of different
models and their calibration according to the data reported by \citet{trel75book}
was carried out by \citet{steinmann&al2012}.

The curves with the square, hollow and filled circle marks in Fig.
(\ref{fig:pure_shear}b) correspond to the macroscopic model, the
Gaussian model and the microscopic model, respectively. Here, the
dependency of $\dimDfield$ on $\dimEfield$ is illustrated. Due to
the constant permittivity, we again note the linear dependency predicted
by the macroscopic model. The Gaussian and the microscopic models,
which are based on the electric long-chains model, predict a change
in the permittivity as a result of the mechanical stretch, in agreement
with the experimental findings of \citet{choi2005effects,wissler2007electromechanical,mcka,qiang2012experimental}.
Since the deformation is dictated by the boundary condition, the Gaussian
and the microscopic models predict the same electric behavior. 

\vspace{7 mm}
\end{onehalfspace}

\begin{onehalfspace}

\subsection{Equibiaxial actuation normal to the electric field\label{sub:Trac Free}}
\end{onehalfspace}

\begin{onehalfspace}
\vspace{7 mm}We once again examine a thin layer of a polymer whose
opposite faces are covered with flexible electrodes with negligible
stiffness. This time, however, the circumferential boundary of the
layer is traction free, thus allowing the layer to expand in the plane
transverse to the direction of the electric field in response to the
electric excitation. We choose the same system of axes as in subsection
\ref{sub:Biaxial} and, thanks to the symmetry of the loading, the
deformation gradient is diagonal with $\stretch_{x}=\stretch_{z}=\stretch$.
Due to the assumed incompressibility we have $\stretch_{y}=\frac{1}{\stretch^{2}}$.

\begin{figure}
\hspace*{\fill} \includegraphics[scale=0.45]{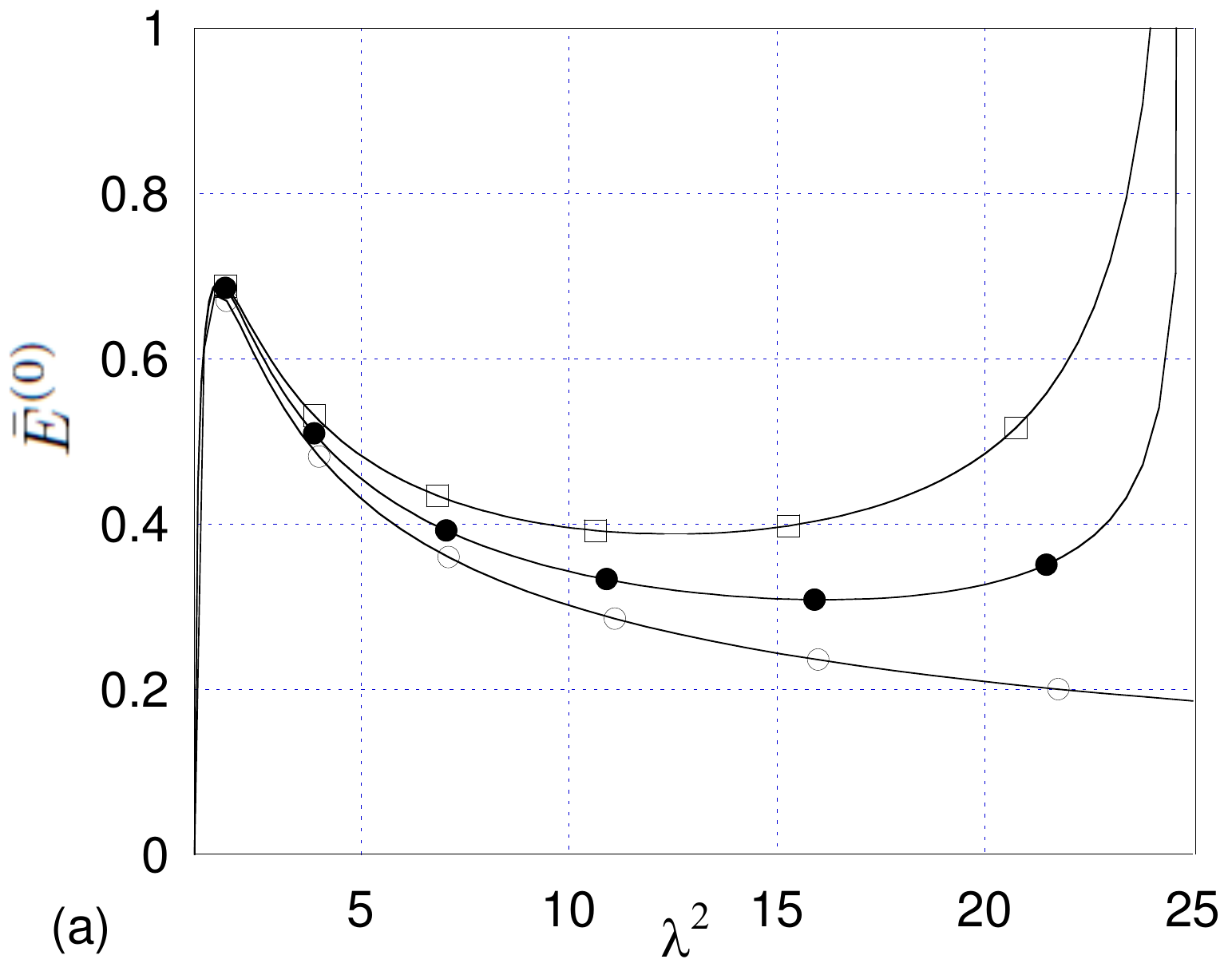}\qquad{}\includegraphics[scale=0.45]{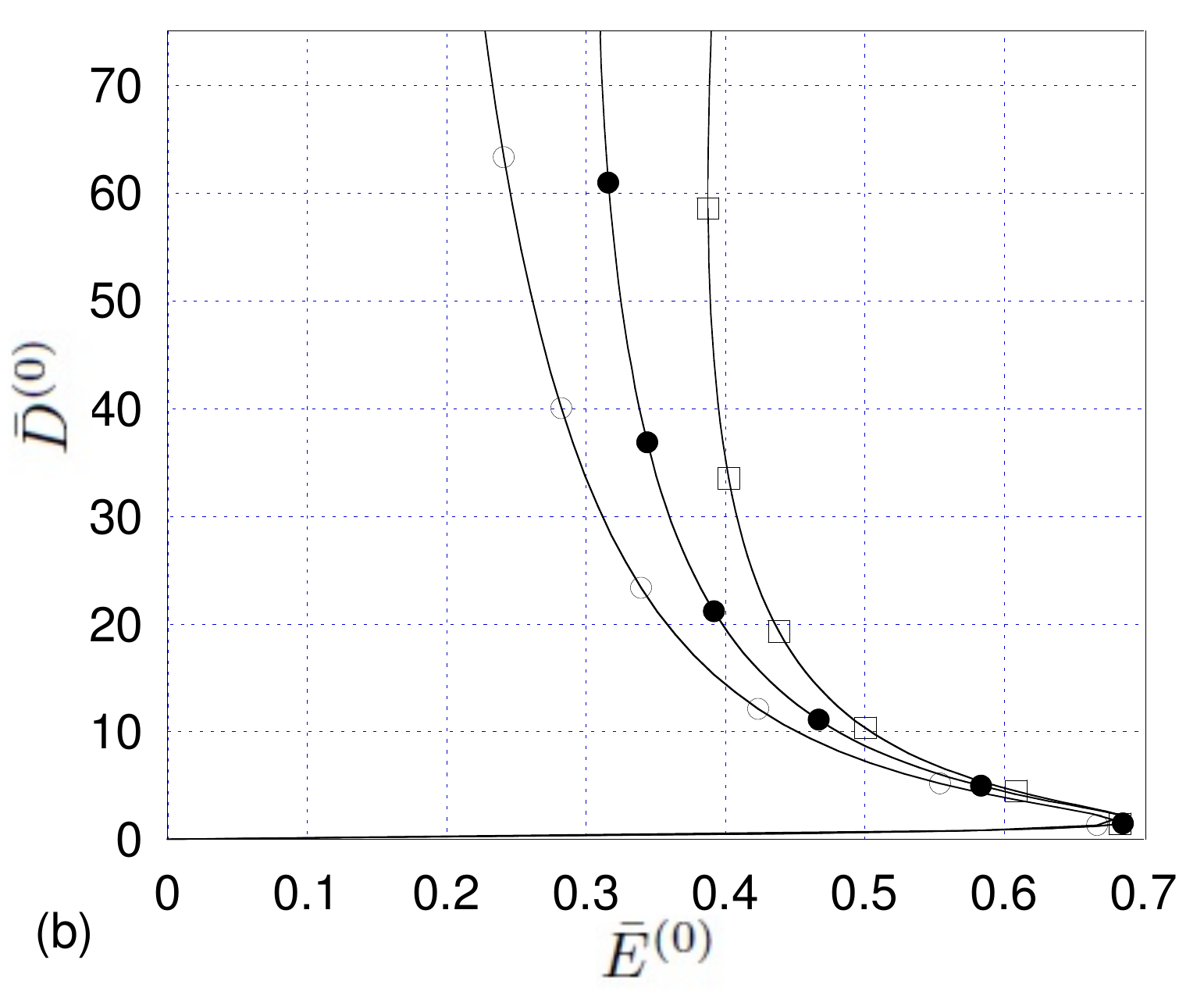}\hspace*{\fill}

\protect\caption{The dimensionless referential electric field $\protect\dimEref$ (a)
and electric displacement $\protect\dimDref$ (b) versus the stretch
of the transverse plane and the dimensionless referential electric
field according to the macroscopic model (the curve with the square
marks), the Gaussian model (the curve with the hollow circle marks)
and the microscopic model (the curve with the filled circle marks).
\label{fig:Traction_free}}
\end{figure}

Fig. (\ref{fig:Traction_free}a) displays the dimensionless referential
electric field $\dimEref$ as a function of the induced stretch of
the transverse plane according to the macroscopic model (the curve
with the square marks), the Gaussian model (the curve with the hollow
circle marks) and the microscopic model (the curve with the filled
circle marks). The loss of stability discussed in the works of \citet{Dorfmann20101,Bertoldi201118,raey}
and \citet{Shmuel2012307} is demonstrated again. We note that after
the peak at $\dimEref\approx0.7$, even though the current electric
field increases monotonically, the Gaussian model predicts a decrease
in the referential electric field with an increase of the stretch.
In an experiment where the referential electric field is controlled,
the macroscopic and microscopic models predict a jump in the planar
stretch. This effect of a transition between two states is known as
snap-through (\citealp{goulbourne2005nonlinear}; \citealp{rudy&etal12ijnm}). 

The curves with the square, hollow and filled circle marks in Fig.
(\ref{fig:Traction_free}b) correspond to the macroscopic, the Gaussian
and the microscopic models, where the predicted dependencies of $\dimDref$
on $\dimEref$ in the direction of the electric field are depicted.
Essentially, this plot illustrates the amount of charge per unit referential
surface area as a function of the potential difference divided by
the initial thickness of the layer. Initially, we observe an increase
of the surface charge with an increase of the electric potential.
However, beyond the peak at $\dimEref\approx0.7$ there is a reversed
trend where, at equilibrium, the electric potential drops while the
surface charge increases. This occurs in conjunction with the uncontrollable
increase in the area of the actuator as shown in Fig. (\ref{fig:Traction_free}a).
From a practical viewpoint this implies that beyond the peak, in a
manner reminiscent of an electrical short-circuit, excessive current
flows from the system electric source while the electric potential
drops. We stress that due to the thinning of the layer the current
electric field increases and may result in a failure of the DE due
to electric breakdown. We also note that even though the same electric
model is used in both the Gaussian and the microscopic models, there
is a difference in the relations between the electric field and the
electric displacement. This is due to the different mechanical deformations
resulting from the applied electric field according to the two different
models.

\vspace{7 mm}
\end{onehalfspace}

\begin{onehalfspace}

\subsection{Uniaxial actuation normal to the electric field \label{sub:Uniaxial-actuation-normal} }
\end{onehalfspace}

\begin{onehalfspace}
\vspace{7 mm}We consider a setting reminiscent of the one considered
in subsection \ref{sub:Trac Free}, but this time the layer is free
to expand only along the $\mathbf{\hat{x}}$-direction. Consequently,
the deformation gradient components are $\stretch_{z}=1$, $\stretch_{x}=\stretch$,
and $\stretch_{y}=\frac{1}{\stretch}$.

\begin{figure}
\hspace*{\fill}\includegraphics[scale=0.45]{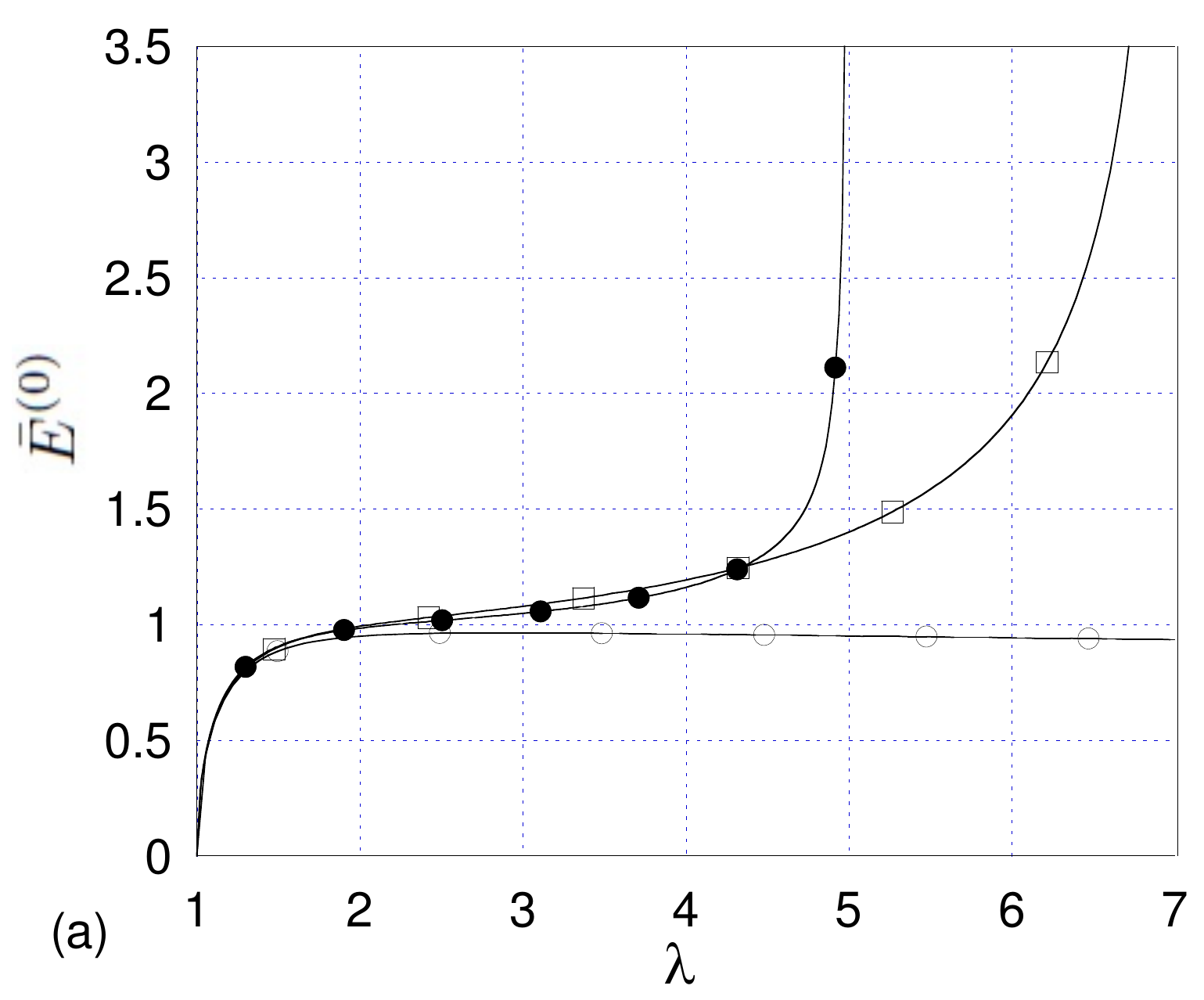}\qquad{}
\includegraphics[scale=0.45]{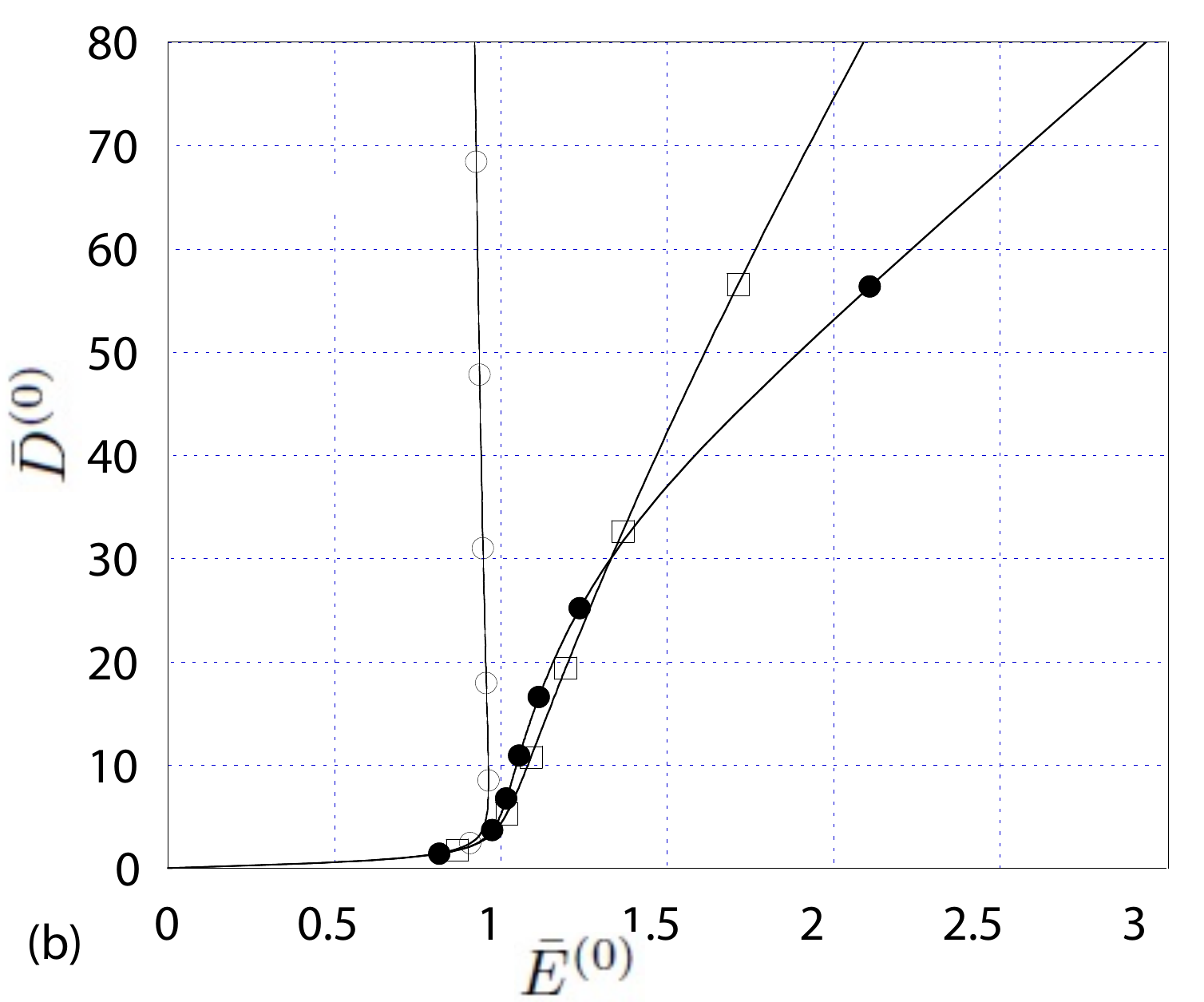}\hspace*{\fill}

\protect\caption{The dimensionless referential electric field $\protect\dimEref$ (a)
and electric displacement $\protect\dimDref$ (b) versus the stretch
and the dimensionless referential electric field according to the
macroscopic model (the curve with the square marks), the Gaussian
model (the curve with the hollow circle marks) and the microscopic
model (the curve with the filled circle marks). \label{fig:actuation_strain}}
\end{figure}

Fig. (\ref{fig:actuation_strain}a) shows the dimensionless referential
electric field as a function of the stretch according to the macroscopic
model (the curve with the square marks), the Gaussian model (the curve
with the hollow circle marks) and the microscopic model (the curve
with the filled circle marks). As mentioned previously, the predicted
lock up stretches are $\stretch^{lu}=5$ and $\stretch^{lu}\approx7$
according to the microscopic and the macroscopic models, respectively.
Unlike the biaxial case described in subsection \ref{sub:Trac Free},
this loading does not admit loss of stability according to the macroscopic
and the microscopic models. The Gaussian model, that is based on the
mechanical neo-Hookean model, reaches a peak at $\dimEref\approx1$
and then monotonically decreases.

Fig. (\ref{fig:actuation_strain}b) depicts the dependence of $\dimDref$
on $\dimEref$. According to the Gaussian model, represented by the
curve with the hollow circle marks, no significant changes in the
surface charge are observed as we initially increase the electric
potential difference. However, beyond the peak of $\dimEref\approx1$,
this model predicts an unstable behavior as a result of the electrical
long-chains model. In order to maintain equilibrium according to the
macroscopic and the microscopic models an increase in the charge on
the electrodes requires an increase in the voltage difference between
them. Thus, the Gaussian model is admitting a behavior that is qualitatively
different from the behaviors of the other two models.

\vspace{7 mm}
\end{onehalfspace}

\begin{onehalfspace}

\section{Concluding remarks}
\end{onehalfspace}

\begin{onehalfspace}
\vspace{7 mm}We determined the behavior of an incompressible polymer
undergoing homogenous deformations according to three electromechanical
models under four types of boundary conditions. The first model incorporates
well-known macroscopically motivated constitutive relations for the
mechanical and the electrical behaviors. The second, microscopic model,
combines mechanical and electrical models stemming from the microstructure
of the polymer. The third model assumes a Gaussian distribution of
the polymer chains and accordingly the mechanical and the electrical
behaviors are determined. We comment that the material parameters
$\numberofdipoles$ and $\numberoflinks$, which denote the number
of dipoles and links, respectively, are used as fitting parameters
and therefore the electrical long-chains model and the Langevin model
are not consistent. Further investigations in this regard is needed. 

In the first two representative examples we apply a referential electric
field by setting the potential difference between the electrodes and
controlling the deformation. In the following two examples we apply
a referential electric field and set the traction on the boundaries
of the polymer. In order to determine the polarization and the stress
resulting according to the microscopic models we make use of the micro-sphere
technique. A comparison between the results shows that the macroscopically
and the microscopically motivated models predict different behaviors.
Therefore, this work encourages a further and a more rigorous investigation
into the connection between the two scales aimed at deepening our
understanding of the micro-macro relations and the mechanisms which
control the actuation. Moreover, analysis of this type may open the
path to the design and manufacturing of polymers with microstructures
that enable to improve the electromechanical coupling.\vspace{7 mm}
\end{onehalfspace}

\begin{onehalfspace}

\subsubsection*{Competing interests}
\end{onehalfspace}

\begin{onehalfspace}
\vspace{2 mm}We have no competing interests.\vspace{7 mm}
\end{onehalfspace}

\begin{onehalfspace}

\subsubsection*{Authors' contributions}
\end{onehalfspace}

\begin{onehalfspace}
\vspace{2 mm}All authors carried out the research and the analysis
jointly and N.C. was responsible for the programming. All authors
gave final approval for publication.\vspace{7 mm}
\end{onehalfspace}

\begin{onehalfspace}

\subsubsection*{Funding }
\end{onehalfspace}

\begin{onehalfspace}
\vspace{2 mm}The first author would like to thank the financial assistance
of the Minerva Foundation. Additionally, partial financial support
for this work was provided by the Swedish Research Council (Vetenskapsr�det)
under grant 2011-5428, and is gratefully acknowledged by the second
author. Lastly, the first and the last authors wish to acknowledge
the support of the Israel Science Foundation founded by the Israel
Academy of Sciences and Humanities (grant 1246/11).

\bibliographystyle{unsrtnat}
\bibliography{bibfile}

\begin{thebibliography}{58}
\providecommand{\natexlab}[1]{#1}
\providecommand{\url}[1]{\texttt{#1}}
\expandafter\ifx\csname urlstyle\endcsname\relax
  \providecommand{\doi}[1]{doi: #1}\else
  \providecommand{\doi}{doi: \begingroup \urlstyle{rm}\Url}\fi

\bibitem[Bar-Cohen(2001)]{barc01book}
Y.~Bar-Cohen.
\newblock {EAP} history, current status, and infrastructure.
\newblock In Y.~Bar-Cohen, editor, \emph{Electroactive Polymer (EAP) Actuators
  as Artificial Muscles}, chapter~1, pages 3--44. SPIE press, Bellingham, WA,
  2001.

\bibitem[McKay et~al.(2010)McKay, O'Brien, Calius, and
  Anderson]{mcka&etal10apl}
T.~McKay, B.~O'Brien, E.~Calius, and L.~Anderson.
\newblock An integrated, self-priming dielectric elastomer generator.
\newblock \emph{Applied Physics Letters}, 97:\penalty0 062911, 2010.

\bibitem[Springhetti et~al.(2014)Springhetti, Bortot, deBotton, and
  Gei]{Springhetti01102014}
R.~Springhetti, E.~Bortot, G.~deBotton, and M.~Gei.
\newblock Optimal energy-harvesting cycles for load-driven dielectric
  generators in plane strain.
\newblock \emph{IMA Journal of Applied Mathematics}, 79\penalty0 (5):\penalty0
  929--946, 2014.

\bibitem[Rudykh et~al.(2012)Rudykh, Bhattacharya, and
  de{B}otton]{rudy&etal12ijnm}
S.~Rudykh, K.~Bhattacharya, and G.~de{B}otton.
\newblock Snap-through actuation of thick-wall electroactive balloons.
\newblock \emph{International Journal of Non-linear Mechanics}, 47:\penalty0
  206--209, 2012.

\bibitem[Gei et~al.(2013)Gei, Springhetti, and Bortot]{0964-1726-22-10-104014}
M.~Gei, R.~Springhetti, and E.~Bortot.
\newblock Performance of soft dielectric laminated composites.
\newblock \emph{Smart Materials and Structures}, 22\penalty0 (10):\penalty0
  104014, 2013.

\bibitem[Shmuel et~al.(2012)Shmuel, Gei, and deBotton]{Shmuel2012307}
G.~Shmuel, M.~Gei, and G.~deBotton.
\newblock The rayleigh-lamb wave propagation in dielectric elastomer layers
  subjected to large deformations.
\newblock \emph{International Journal of Non-Linear Mechanics}, 47\penalty0
  (2):\penalty0 307 -- 316, 2012.
\newblock Nonlinear Continuum Theories.

\bibitem[Toupin(1956)]{toup56arma}
R.~A. Toupin.
\newblock The elastic dielectric.
\newblock \emph{Journal of Rational Mechanics and Analysis}, 5:\penalty0
  849--915, 1956.

\bibitem[Kofod et~al.(2003)Kofod, Sommer-Larsen, Kornbluh, and
  Pelrine]{kofo&etal03jimss}
G.~Kofod, P.~Sommer-Larsen, R.~Kornbluh, and R.~Pelrine.
\newblock Actuation response of polyacrylate dielectric elastomers.
\newblock \emph{Journal of Intelligent Material Systems and Structures},
  14:\penalty0 787--793, 2003.

\bibitem[Pelrine et~al.(2000{\natexlab{a}})Pelrine, Kornbluh, Joseph, Heydt,
  Pei, and Chiba]{Pelrine200089}
R.~Pelrine, R.~Kornbluh, J.~Joseph, R.~Heydt, Q.~Pei, and S.~Chiba.
\newblock High-field deformation of elastomeric dielectrics for actuators.
\newblock \emph{Materials Science and Engineering: C}, 11\penalty0
  (2):\penalty0 89 -- 100, 2000{\natexlab{a}}.

\bibitem[Pelrine et~al.(2000{\natexlab{b}})Pelrine, Kornbluh, Pei, and
  Joseph]{Pelrine04022000}
R.~Pelrine, R.~Kornbluh, Q.~Pei, and J.~Joseph.
\newblock High-speed electrically actuated elastomers with strain greater than
  100\%.
\newblock \emph{Science}, 287\penalty0 (5454):\penalty0 836--839,
  2000{\natexlab{b}}.

\bibitem[Huang et~al.(2004)Huang, Zhang, de{B}otton, and
  Bhattacharya]{huan&etal04aple}
C.~Huang, Q.~M. Zhang, G.~de{B}otton, and K.~Bhattacharya.
\newblock All-organic dielectric-percolative three-component composite
  materials with high electromechanical response.
\newblock \emph{Applied Physics Letters}, 84:\penalty0 4391--4393, 2004.

\bibitem[Stoyanov et~al.(2010)Stoyanov, Kollosche, McCarthy, and
  Kofod]{stoy&etal10jmatchem}
H.~Stoyanov, M.~Kollosche, D.~N. McCarthy, and G.~Kofod.
\newblock Molecular composites with enhanced energy density for electroactive
  polymers.
\newblock \emph{Journal of Materials Chemistry}, 20:\penalty0 7558--7564, 2010.

\bibitem[Tian et~al.(2012)Tian, Tevet-Deree, de{B}otton, and
  Bhattacharya]{tian12jmps}
L.~Tian, L.~Tevet-Deree, G.~de{B}otton, and K.~Bhattacharya.
\newblock Dielectric elastomer composites.
\newblock \emph{Journal of the Mechanics and Physics of Solids}, 60\penalty0
  (1):\penalty0 181 -- 198, 2012.

\bibitem[Galipeau and Casta{\~n}eda(2012)]{Galipeau20121}
E.~Galipeau and P.~Ponte Casta{\~n}eda.
\newblock The effect of particle shape and distribution on the macroscopic
  behavior of magnetoelastic composites.
\newblock \emph{International Journal of Solids and Structures}, 49\penalty0
  (1):\penalty0 1 -- 17, 2012.

\bibitem[Rudykh et~al.(2013)Rudykh, Lewinstein, Uner, and
  deBotton]{:/content/aip/journal/apl/102/15/10.1063/1.4801775}
S.~Rudykh, A.~Lewinstein, G.~Uner, and G.~deBotton.
\newblock Analysis of microstructural induced enhancement of electromechanical
  coupling in soft dielectrics.
\newblock \emph{Applied Physics Letters}, 102\penalty0 (15):\penalty0 --, 2013.

\bibitem[Lopez-Pamies(2014)]{lopez2014elastic}
O.~Lopez-Pamies.
\newblock Elastic dielectric composites: Theory and application to
  particle-filled ideal dielectrics.
\newblock \emph{Journal of the Mechanics and Physics of Solids}, 64:\penalty0
  61--82, 2014.

\bibitem[Ba{\v{z}}ant and Oh(1986)]{bavzant1986efficient}
P.~Ba{\v{z}}ant and B.H. Oh.
\newblock Efficient numerical integration on the surface of a sphere.
\newblock \emph{ZAMM-Journal of Applied Mathematics and Mechanics/Zeitschrift
  f{\"u}r Angewandte Mathematik und Mechanik}, 66\penalty0 (1):\penalty0
  37--49, 1986.

\bibitem[Carol et~al.(2004)Carol, Jirasek, and Ba{\v{z}}ant]{Carol2004511}
I.~Carol, M.~Jirasek, and Z.~P. Ba{\v{z}}ant.
\newblock A framework for microplane models at large strain, with application
  to hyperelasticity.
\newblock \emph{International Journal of Solids and Structures}, 41\penalty0
  (2):\penalty0 511 -- 557, 2004.

\bibitem[Miehe et~al.(2004)Miehe, G\"oktepe, and Lulei]{Miehe20042617}
C.~Miehe, S.~G\"oktepe, and F.~Lulei.
\newblock A micro-macro approach to rubber-like materials--part i: the
  non-affine micro-sphere model of rubber elasticity.
\newblock \emph{Journal of the Mechanics and Physics of Solids}, 52\penalty0
  (11):\penalty0 2617 -- 2660, 2004.

\bibitem[Thylander et~al.(2012)Thylander, Menzel, and
  Ristinmaa]{0964-1726-21-9-094008}
S.~Thylander, A.~Menzel, and M.~Ristinmaa.
\newblock An electromechanically coupled micro-sphere framework: application to
  the finite element analysis of electrostrictive polymers.
\newblock \emph{Smart Materials and Structures}, 21\penalty0 (9):\penalty0
  094008, 2012.

\bibitem[Ogden(1997)]{ogden97book}
R.~W. Ogden.
\newblock \emph{Non-Linear Elastic Deformations}.
\newblock Dover Publications, New York, 1997.

\bibitem[Kuhn and Gr{\"u}n(1942)]{kuhn1942beziehungen}
W.~Kuhn and F.~Gr{\"u}n.
\newblock Beziehungen zwischen elastischen konstanten und
  dehnungsdoppelbrechung hochelastischer stoffe.
\newblock \emph{Kolloid-Zeitschrift}, 101\penalty0 (3):\penalty0 248--271,
  1942.

\bibitem[Wang and Guth(1952)]{wang&guth52jcp}
M.~C. Wang and E.~Guth.
\newblock Statistical theory of networks of non-gaussian flexible chains.
\newblock \emph{The Journal of Chemical Physics}, 20:\penalty0 1144--1157,
  1952.

\bibitem[Flory and
  Rehner(1943)]{:/content/aip/journal/jcp/11/11/10.1063/1.1723791}
P.~J. Flory and J.~Rehner.
\newblock Statistical mechanics of cross-linked polymer networks i. rubberlike
  elasticity.
\newblock \emph{The Journal of Chemical Physics}, 11\penalty0 (11):\penalty0
  512--520, 1943.

\bibitem[Treloar(1946)]{TF9464200083}
L.~R.~G. Treloar.
\newblock The elasticity of a network of long-chain molecules.-{III}.
\newblock \emph{Transactions of the Faraday Society}, 42:\penalty0 83--94,
  1946.

\bibitem[Arruda and Boyce(1993)]{arru&boyc93jmps}
E.~M. Arruda and M.~C. Boyce.
\newblock A three-dimensional constitutive model for the large stretch behavior
  of rubber elastic materials.
\newblock \emph{Journal of the Mechanics and Physics of Solids}, 41:\penalty0
  389--412, 1993.

\bibitem[Tiersten(1990)]{tier90book}
H.~F. Tiersten.
\newblock \emph{A Development of the Equations of Electromagnetism in Material
  Continua}, volume~36 of \emph{Springer Tracts in Natural Philosophy}.
\newblock Springer-Verlag, New York, 1990.

\bibitem[Hutter et~al.(2006)Hutter, van~de Ven, and Ursescu]{hutt&etal06book}
K.~Hutter, A.~A.~F. van~de Ven, and A.~Ursescu.
\newblock \emph{Electromagnetic Field Matter Interactions in Thermoelasic
  Solids and Viscous Fluids}.
\newblock Number 710 in Lecture Notes in Physics. Springer, Berlin Heidelberg,
  2nd edition, 2006.

\bibitem[Dorfmann and Ogden(2005)]{dorf&ogde05acmc}
A.~Dorfmann and R.~W. Ogden.
\newblock Nonlinear electroelasticity.
\newblock \emph{Acta Mechanica}, 174:\penalty0 167--183, 2005.

\bibitem[Ask et~al.(2012{\natexlab{a}})Ask, Menzel, and Ristinmaa]{Ask2012156}
A.~Ask, A.~Menzel, and M.~Ristinmaa.
\newblock Phenomenological modeling of viscous electrostrictive polymers.
\newblock \emph{International Journal of Non-Linear Mechanics}, 47\penalty0
  (2):\penalty0 156 -- 165, 2012{\natexlab{a}}.

\bibitem[Ask et~al.(2012{\natexlab{b}})Ask, Menzel, and Ristinmaa]{Ask20129}
A.~Ask, A.~Menzel, and M.~Ristinmaa.
\newblock Electrostriction in electro-viscoelastic polymers.
\newblock \emph{Mechanics of Materials}, 50\penalty0 (0):\penalty0 9 -- 21,
  2012{\natexlab{b}}.

\bibitem[Jim{\'e}nez and McMeeking(2013)]{jimenez2013deformation}
S.~Jim{\'e}nez and R.~M. McMeeking.
\newblock Deformation dependent dielectric permittivity and its effect on
  actuator performance and stability.
\newblock \emph{International Journal of Non-Linear Mechanics}, 57:\penalty0
  183--191, 2013.

\bibitem[Cohen and deBotton(2014{\natexlab{a}})]{Cohen14b}
N.~Cohen and G.~deBotton.
\newblock The electromechanical response of polymer networks with long-chain
  molecules.
\newblock \emph{Mathematics and Mechanics of Solids}, 2014{\natexlab{a}}.
\newblock \doi{10.1177/1081286514550574}.

\bibitem[Cohen and deBotton(2014{\natexlab{b}})]{Cohen2014}
N.~Cohen and G.~deBotton.
\newblock Multiscale analysis of the electromechanical coupling in dielectric
  elastomers.
\newblock \emph{European Journal of Mechanics - A/Solids}, 48\penalty0
  (0):\penalty0 48 -- 59, 2014{\natexlab{b}}.

\bibitem[Mc{M}eeking and Landis(2005)]{mcme&land05jamt}
R.~M. Mc{M}eeking and C.~M. Landis.
\newblock Electrostatic forces and stored energy for deformable dielectric
  materials.
\newblock \emph{Journal of Applied Mechanics, Transactions ASME}, 72:\penalty0
  581--590, 2005.

\bibitem[McMeeking et~al.(2007)McMeeking, Landis, and Jimenez]{mcme&etal07ijnm}
R.~M. McMeeking, C.~M. Landis, and S.~M.~A. Jimenez.
\newblock A principle of virtual work for combined electrostatic and mechanical
  loading of materials.
\newblock \emph{International Journal of Nonlinear Mechanics}, 42\penalty0
  (6):\penalty0 831--838, 2007.

\bibitem[Gent(1996)]{gent96rc&t}
A.~N. Gent.
\newblock A new constitutive relation for rubber.
\newblock \emph{Rubber Chemistry and Technology}, 69:\penalty0 59--61, 1996.

\bibitem[Blythe and Bloor(2008)]{blyt&bloo08book}
T.~Blythe and D.~Bloor.
\newblock \emph{Electrical Properties of Polymers}.
\newblock Cambridge University Press, Cambridge, UK, 2 edition, 2008.

\bibitem[Di~Lillo et~al.(2012)Di~Lillo, Schmidt, Carnelli, Ermanni, Kovacs,
  Mazza, and Bergamini]{:/content/aip/journal/jap/111/2/10.1063/1.3676201}
L.~Di~Lillo, A.~Schmidt, D.~A. Carnelli, P.~Ermanni, G.~Kovacs, E.~Mazza, and
  A.~Bergamini.
\newblock Measurement of insulating and dielectric properties of acrylic
  elastomer membranes at high electric fields.
\newblock \emph{Journal of Applied Physics}, 111\penalty0 (2):\penalty0 --,
  2012.

\bibitem[Choi et~al.(2005)Choi, Jung, Chuc, Jung, Koo, Koo, Lee, Lee, Nam, Cho,
  and Lee]{choi2005effects}
H.~R. Choi, K.~Jung, N.~H. Chuc, M.~Jung, I.~Koo, J.~Koo, J.~Lee, J.~Lee,
  J.~Nam, M.~Cho, and Y.~Lee.
\newblock Effects of prestrain on behavior of dielectric elastomer actuator.
\newblock In \emph{Smart Structures and Materials}, pages 283--291.
  International Society for Optics and Photonics, 2005.

\bibitem[Wissler and Mazza(2007)]{wissler2007electromechanical}
M.~Wissler and E.~Mazza.
\newblock Electromechanical coupling in dielectric elastomer actuators.
\newblock \emph{Sensors and Actuators A: Physical}, 138\penalty0 (2):\penalty0
  384--393, 2007.

\bibitem[McKay et~al.(2009)McKay, Calius, and Anderson]{mcka}
T.~G. McKay, E.~Calius, and I.~A. Anderson.
\newblock The dielectric constant of 3m vhb: a parameter in dispute, 2009.

\bibitem[Qiang et~al.(2012)Qiang, Chen, and Li]{qiang2012experimental}
J.~Qiang, H.~Chen, and B.~Li.
\newblock Experimental study on the dielectric properties of polyacrylate
  dielectric elastomer.
\newblock \emph{Smart Materials and Structures}, 21\penalty0 (2):\penalty0
  025006, 2012.

\bibitem[Flory(1953)]{flor53book}
P.~J. Flory.
\newblock \emph{Principles of polymer chemistry}.
\newblock Cornell Univ Press, Ithaca, NY, 1953.

\bibitem[Treloar(1943)]{trel43atfaradaysoc}
L.~R.~G. Treloar.
\newblock The elasticity of a network of long-chain molecules. {I}.
\newblock \emph{Transactions of the Faraday Society}, 39:\penalty0 36--41,
  1943.

\bibitem[Treloar(1975)]{trel75book}
L.~R.~G. Treloar.
\newblock \emph{The Physics of Rubber Elasticity}.
\newblock Clarendon Press, Oxford, 1975.

\bibitem[Kuhl et~al.(2006)Kuhl, Menzel, and
  Garikipati]{doi:10.1080/14786430500080296}
E.~Kuhl, A.~Menzel, and K.~Garikipati.
\newblock On the convexity of transversely isotropic chain network models.
\newblock \emph{Philosophical Magazine}, 86\penalty0 (21-22):\penalty0
  3241--3258, 2006.

\bibitem[Stockmayer(1967)]{stockmayer1967dielectric}
W.~H. Stockmayer.
\newblock Dielectric dispersion in solutions of flexible polymers.
\newblock \emph{Pure and Applied Chemistry}, 15\penalty0 (539):\penalty0 2816,
  1967.

\bibitem[Menzel and Waffenschmidt(2009)]{menzel2009microsphere}
A.~Menzel and T.~Waffenschmidt.
\newblock A microsphere-based remodelling formulation for anisotropic
  biological tissues.
\newblock \emph{Philosophical Transactions of the Royal Society A:
  Mathematical, Physical and Engineering Sciences}, 367\penalty0
  (1902):\penalty0 3499--3523, 2009.

\bibitem[Alastru{\'e} et~al.(2009{\natexlab{a}})Alastru{\'e}, Mart{\'i}­nez,
  Doblar{\'e}, and Menzel]{Alastru2009178}
V.~Alastru{\'e}, M.A. Mart{\'i}­nez, M.~Doblar{\'e}, and A.~Menzel.
\newblock Anisotropic micro-sphere-based finite elasticity applied to blood
  vessel modelling.
\newblock \emph{Journal of the Mechanics and Physics of Solids}, 57\penalty0
  (1):\penalty0 178 -- 203, 2009{\natexlab{a}}.

\bibitem[Alastru{\'e} et~al.(2009{\natexlab{b}})Alastru{\'e}, Mart{\'i}­nez,
  Menzel, and Doblar{\'e}]{NME:NME2577}
V.~Alastru{\'e}, M.A. Mart{\'i}­nez, A.~Menzel, and M.~Doblar{\'e}.
\newblock On the use of non-linear transformations for the evaluation of
  anisotropic rotationally symmetric directional integrals. application to the
  stress analysis in fibred soft tissues.
\newblock \emph{International Journal for Numerical Methods in Engineering},
  79\penalty0 (4):\penalty0 474--504, 2009{\natexlab{b}}.

\bibitem[Waffenschmidt et~al.(2012)Waffenschmidt, Menzel, and
  Kuhl]{Waffenschmidt20121928}
T.~Waffenschmidt, A.~Menzel, and E.~Kuhl.
\newblock Anisotropic density growth of bone-a computational micro-sphere
  approach.
\newblock \emph{International Journal of Solids and Structures}, 49\penalty0
  (14):\penalty0 1928 -- 1946, 2012.

\bibitem[Ostwald et~al.(2014)Ostwald, Bartel, and Menzel]{NME:NME4601}
R.~Ostwald, T.~Bartel, and A.~Menzel.
\newblock A gibbs-energy-barrier-based computational micro-sphere model for the
  simulation of martensitic phase-transformations.
\newblock \emph{International Journal for Numerical Methods in Engineering},
  97\penalty0 (12):\penalty0 851--877, 2014.

\bibitem[Dorfmann and Ogden(2010)]{Dorfmann20101}
A.~Dorfmann and R.W. Ogden.
\newblock Nonlinear electroelastostatics: Incremental equations and stability.
\newblock \emph{International Journal of Engineering Science}, 48\penalty0
  (1):\penalty0 1 -- 14, 2010.

\bibitem[Bertoldi and Gei(2011)]{Bertoldi201118}
K.~Bertoldi and M.~Gei.
\newblock Instabilities in multilayered soft dielectrics.
\newblock \emph{Journal of the Mechanics and Physics of Solids}, 59\penalty0
  (1):\penalty0 18 -- 42, 2011.

\bibitem[Steinmann et~al.(2012)Steinmann, Hossain, and
  Possart]{steinmann&al2012}
P.~Steinmann, M.~Hossain, and G.~Possart.
\newblock Hyperelastic models for rubber-like materials: consistent tangent
  operators and suitability for treloar's data.
\newblock \emph{Archive of Applied Mechanics}, 82\penalty0 (9):\penalty0
  1183--1217, 2012.

\bibitem[Rudykh and deBotton(2011)]{raey}
S.~Rudykh and G.~deBotton.
\newblock Stability of anisotropic electroactive polymers with application to
  layered media.
\newblock \emph{Zeitschrift f{\"u}r angewandte Mathematik und Physik},
  62\penalty0 (6):\penalty0 1131--1142, 2011.

\bibitem[Goulbourne et~al.(2005)Goulbourne, Mockensturm, and
  Frecker]{goulbourne2005nonlinear}
N.~Goulbourne, E.~Mockensturm, and M.~Frecker.
\newblock A nonlinear model for dielectric elastomer membranes.
\newblock \emph{Journal of Applied Mechanics}, 72\penalty0 (6):\penalty0
  899--906, 2005.

\end{thebibliography}
\end{onehalfspace}

\end{document}